\documentclass[journal=cmatex,layout=draft,email=false]{achemso}
\setkeys{acs}{
    articletitle = true,
}
\setcitestyle{square}
\setkeys{acs}{etalmode=truncate,maxauthors=0}
\usepackage[version=4]{mhchem}
\usepackage[hyperindex]{hyperref}
\hypersetup{colorlinks,citecolor=blue,linkcolor=blue,urlcolor=blue}
\usepackage{units}
\usepackage{amsmath}
\usepackage{booktabs}

\newcommand{\diffd}{\,\text{d}}
\newcommand{\voc}{V_\text{oc}}
\newcommand{\vint}{V_\text{int}}
\newcommand{\jsc}{j_\text{sc}}
\newcommand{\nid}{n_\text{id}}
\newcommand{\tex}{t_\text{ex}}
\newcommand{\trec}{t_\text{rec}}
\newcommand{\dq}{\Delta Q}
\newcommand{\kb}{k_\text{B}}
\newcommand{\kenc}{k_\text{enc}}
\newcommand{\zetaCT}{\zeta_\text{CT}}

\newcommand{\etaCT}{\eta_\text{CT,diss}}

\author{Sebastian Wilken}
\affiliation{Physics, Faculty of Science and Engineering, \AA{}bo Akademi University, Porthansgatan 3, 20500 Turku, Finland}
\alsoaffiliation{Institute of Physics, Carl von Ossietzky University of Oldenburg, 26111 Oldenburg, Germany}
\email{sebastian.wilken@abo.fi}

\author{Dorothea Scheunemann}
\affiliation{Department of Chemistry and Chemical Engineering, Chalmers University of Technology, 412 96 G\"oteborg, Sweden}
\alsoaffiliation{Institute of Physics, Carl von Ossietzky University of Oldenburg, 26111 Oldenburg, Germany}

\author{Staffan Dahlstr\"{o}m}
\author{Mathias Nyman}
\affiliation{Physics, Faculty of Science and Engineering, \AA{}bo Akademi University, Porthansgatan 3, 20500 Turku, Finland}

\author{J\"urgen Parisi}
\affiliation{Institute of Physics, Carl von Ossietzky University of Oldenburg, 26111 Oldenburg, Germany}

\author{Ronald \"{O}sterbacka}
\affiliation{Physics, Faculty of Science and Engineering, \AA{}bo Akademi University, Porthansgatan 3, 20500 Turku, Finland}

\title{How to Reduce Charge Recombination in Organic Solar Cells: There Are Still Lessons to Learn from P3HT:PCBM}

\SectionNumbersOn

\begin{document}

\begin{abstract}
Suppressing charge recombination is key for organic solar cells to become commercial reality. However, there is still no conclusive picture of how recombination losses are influenced by the complex nanoscale morphology. Here, new insight is provided by revisiting the P3HT:PCBM~blend, which is still one of the best performers regarding reduced recombination. By changing small details in the annealing procedure, two model morphologies were prepared that vary in phase separation, molecular order and phase purity, as revealed by electron tomography and optical spectroscopy. Both systems behave very similarly with respect to charge generation and transport, but differ significantly in bimolecular recombination. Only the system containing P3HT aggregates of high crystalline quality and purity is found to achieve exceptionally low recombination rates. The high-quality aggregates support charge delocalization, which assists the re-dissociation of interfacial charge-transfer states formed upon the encounter of free carriers. For devices with the optimized morphology, an exceptional long hole diffusion length is found, which allows them to work as Shockley-type solar cells even in thick junctions of \unit[300]{nm}. In contrast, the encounter rate and the size of the phase-separated domains appears to be less important.
\end{abstract}

\section{Introduction}
The emergence of nonfullerene acceptors~(NFAs) has pushed bulk-heterojunction organic photovoltaics~(OPVs) to record efficiencies approaching~20\%.\cite{Firdaus2019,Cui2020,Lin2020} However, despite the wide variety of materials that are now available, there are still only a few systems that maintain their full performance at junction thicknesses of~\unit[300]{nm} and more.\cite{Armin2018,Sun2018,Jin2017} Compatibility with thick active layers, as well as a general tolerance to thickness variations, is considered an important prerequisite for the low-cost production of OPVs using printing techniques.\cite{Meredith2018,Clarke2014} The main problem with increasing thickness is the slowdown of charge collection, which makes photogenerated carriers more vulnerable to recombination.\cite{Bartesaghi2015,Kaienburg2016,Neher2016} This is particularly true for NFA-based systems, which typically have low carrier mobilities of~$10^{-9}$ to $\unit[10^{-8}]{m^2\,V^{-1}s^{-1}}$. Hence, in order to compensate for the limitations in transport, it has become extremely important to find strategies how charge recombination can be suppressed. 

Despite considerable efforts, there is still no agreement on which are the key factors determining the recombination strength.\cite{Goehler2018,Lakhwani2014} Conceptually, free charge recombination is a bimolecular process and can be described by the rate equation~$R = k_2 n^2$, where $k_2$ is the rate constant and $n$ the carrier density. The parameter~$k_2$ is often compared with the homogeneous Langevin model,
\begin{equation}
k_L = \frac{q}{\varepsilon\varepsilon_0} (\mu_n + \mu_p),
\label{eq:Langevin}
\end{equation}
where $q$ is the elementary charge, $\varepsilon\varepsilon_0$ the dielectric permittivity, $\mu_n$ the electron mobility and $\mu_p$ the hole mobility. Although there are a number of systems in which~$k_2$ is significantly reduced compared to Langevin recombination~($k_2 = \zeta k_L$, where $\zeta < 1$), the reduction is usually not great enough to ensure thickness-insensitive device performance. It is generally accepted that the reduction factor~$\zeta$ is affected by the blend morphology, but the details remain controversial. For example, while some authors relate~$\zeta$ to the phase separation between the donor and acceptor,\cite{Pivrikas2005,Koster2006,Groves2008,Heiber2015,Coropceanu2017} others highlight the importance of phase purity and molecular order.\cite{Burke2015,Nyman2015,Cha2019,Schwarz2020,Wilken2020b,Liu2020,Hosseini2020} The difficulties of manipulating the morphology in a controlled way makes experimental clarification a complex task.

Among the benchmark systems for reduced recombination are blends based on the classical polymer poly(3-hexyl\-thiophene)~(P3HT). Both with fullerene and NFAs, reduction factors as low as~$10^{-4}$ have been reported.\cite{Ferguson2011,Pivrikas2005,Gasparini2017} In particular, the availability of NFAs that complement the absorption of P3HT and minimize voltage losses due to a fine-tuned energy level alignment has led to a remarkable renaissance of P3HT in OPV~research.\cite{Khan2019,Holliday2016,Gasparini2017,Baran2017,Xiao2017,Wu2015,Wu2019,Yang2020} One special feature of P3HT-based OPVs is that they develop their full performance only after the application of post-deposition treatments such as thermal and solvent annealing. The annealing induces phase separation and crystallization, thereby transforming the active layer into an ``optimized'' morphology in terms of charge generation and transport.\cite{Kniepert2014,Deibel2008,Li2005} However, a closer look at the literature reveals that the connection between the morphological changes and the recombination is much less clear. In the Supporting Information~(Table~S1 and Figure~S1) we collected 16 studies on annealed blends of P3HT and phenyl-\ce{C61}-butyric acid methyl ester~(PCBM), which is the most studied system to date. Even for nominally equally processed devices, reported reduction factors~$\zeta$ span over 3~orders of magnitude. Although the variation may be partly explained by different measurement methods, it points towards an underlying structure--property relationship that is yet to be revealed. Understanding this relation, or simply answering the question ``What makes P3HT performing so well?'', will help in designing new materials for commercially relevant OPVs.

In this article, we revisit the P3HT:PCBM~blend to determine the key morphological features to suppress charge recombination in OPVs. Specifically, we consider two model morphologies that behave similarly with respect to generation and transport, but show clear differences in recombination strength. By combining electron tomography, optical spectroscopy and electrical measurements, we draw clear connections between morphology and the thickness-dependent competition between charge collection and recombination. We find that the presence of aggregates of high crystalline quality and purity is crucial for OPVs to function as Shockley-type devices without recombination losses even at high thickness. In contrast, the experiments show that the size of the phase-separated domains is of secondary importance. Our results suggest that delocalization along conjugated chain segments is key to allow charge-transfer pairs formed by the encounter of free electrons and holes to re-dissociate rather than to recombine into the ground state. This provides clear design rules for future OPV~materials.

\section{Results}
\subsection{Sample Systems and Device Performance}
\label{sec:systems}

\begin{figure*}[t]
\includegraphics[width=\textwidth]{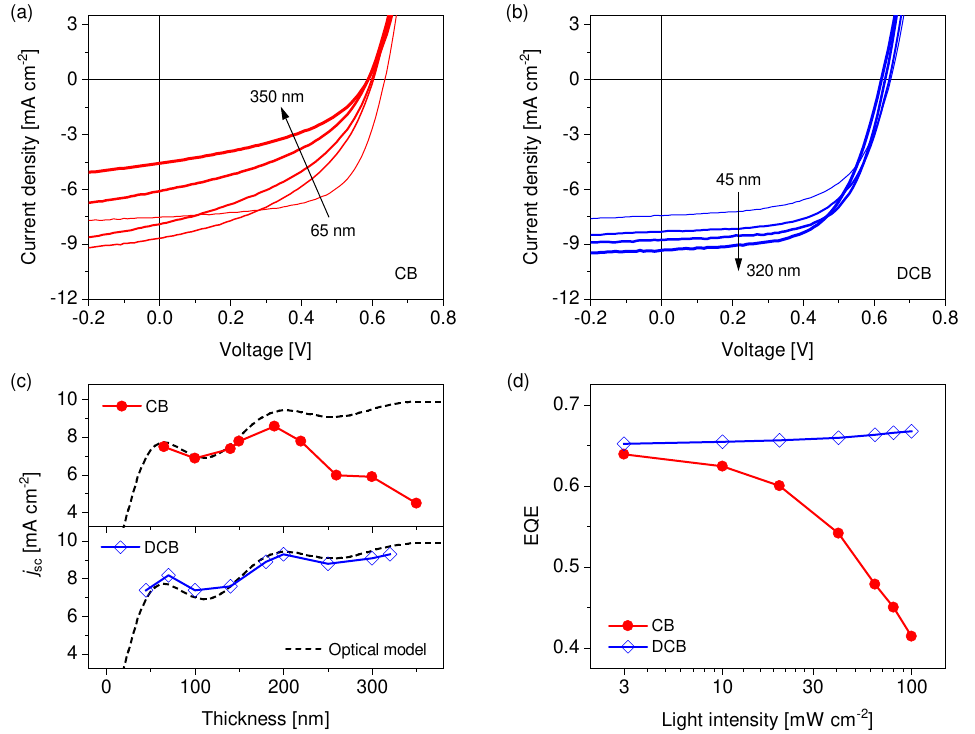}
\caption{Thickness-dependent performance of CB and DCB~devices. (a,b)~$j$--$V$ curves under simulated sunlight for different active-layer thicknesses. See Table~S2 in the Supporting Information for the full data set. (c)~Measured $\jsc$~(symbols) together with the result of transfer-matrix calculations~(dashed lines) assuming a constant IQE of~$70\%$. (d)~White-light bias dependent EQE for 300-nm thick devices.}
\label{fig:figure01}
\end{figure*}

Throughout the following, we compare P3HT:PCBM films in 1:1~blend ratio that were spin-coated from either chlorobenzene~(CB) or 1,2-dichlorobenzene~(DCB) and subsequently thermally annealed. The main effect of the solvent is on the drying rate. While films from CB dried rapidly during spin coating, the ones from DCB were still wet after deposition and let dry slowly before the thermal anneal was applied. We implemented the differently prepared blends into inverted solar cells~(substrate\slash{}cathode\slash{}active layer\slash{}anode) and varied the thickness~$L$ of the active layer. Figure~\ref{fig:figure01}a,b and Table~S2 in the Supporting Information summarize the thickness-dependent current--voltage~($j$--$V$) characteristics under simulated solar illumination. While the performance of the CB and DCB~systems is very similar in thin devices~(efficiency around~3\%), significant differences become apparent with increasing thickness. Clearly, only the slow grown DCB~devices maintain~(and even  improve) their performance by increasing~$L$ from around~50 to over~\unit[300]{nm}. In contrast, for the CB~devices, the photocurrent of becomes increasingly voltage-dependent with increasing thickness, which leads to a clear drop in short-circuit current~($\jsc$) and fill factor~(FF). As a result, the thickest CB~device~($L = \unit[350]{nm}$) achieves an efficiency of only~$1.1\%$.

Figure~\ref{fig:figure01}c shows that the photocurrent of the DCB~devices is well described by an optical transfer-matrix model assuming an internal quantum efficiency~(IQE) independent on thickness. We determined the IQE experimentally using the external quantum efficiency~(EQE) measured with lock-in technique~(see Figure~S3 in the Supporting Information). In the case of the CB~devices, this method fails to explain the drop of $\jsc$ for~$L > \unit[150]{nm}$. The reason is that the EQE decreases strongly with increasing light intensity, which can be seen by adding bias illumination to the low-intensity probe in the EQE measurement~(Figure~\ref{fig:figure01}d). Hence, the poor performance is not an inherent property of the thick CB~devices, but develops gradually with increasing carrier density. This aspect is further elaborated in Figures~S4 and~S5 in the Supporting Information, where light-intensity dependent $j$--$V$~curves and EQE~spectra are shown. Such a behavior is usually attributed to bimolecular recombination, often accompanied by space-charge effects.\cite{Wilken2020} Importantly, at low intensity, all CB and DCB devices display a rather high~IQE of about 70\%. This illustrates that free charge carriers are efficiently generated in both systems, while the differences lie in how the collection of those carriers competes with recombination.\cite{Bartesaghi2015,Neher2016}

Having shown that seemingly small details in the blend preparation have drastic effect on the thickness-dependent device performance, we now want to establish relationships with the morphology. For this purpose, we will first present a detailed characterization of the relevant structural features, that is, phase separation, aggregation and phase purity. We then use mobility measurements, as well as transient photocurrent and photovoltage studies to relate these properties  to the kinetics of charge collection and recombination.

\subsection{Phase Separation}
\label{sec:tem}
\begin{figure*}[t]
\includegraphics[width=\linewidth]{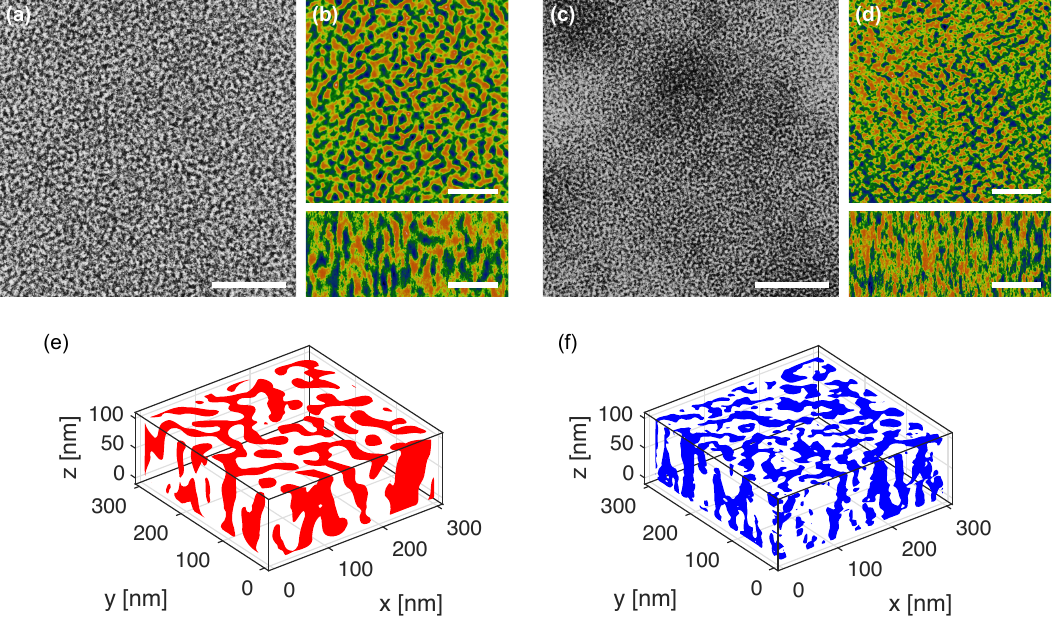}
\caption{Morphology of \unit[200]{nm} thick P3HT:PCBM films processed from CB~(panels a,b and e) and DCB~(panels c,d and f), respectively. (a,c)~Regular bright-field TEM images of free-standing films. Scale bar: \unit[300]{nm}. (b,d)~Exemplary slices through electron tomographic reconstructions parallel~(top) and perpendicular~(bottom) to the film plane. The color coding represents the brightness value of a certain pixel, which decreases from red over green to blue. Scale bar:~\unit[100]{nm}. (e,f)~Representative volume elements, reassembled from the $xy$~slices after binarization. See also Movie~S1 in the Supporting Information.}
\label{fig:figure02}
\end{figure*}

To investigate the phase separation in real space, we used high-resolution transmission electron microscopy~(TEM) and tomography~(Figure~\ref{fig:figure02}). Blend films of $\unit[{\sim}200]{nm}$ thickness were selected for this analysis because they are representative of the thickness series and still exhibit high enough electron transparency.\cite{Oosterhout2009,vanBavel2009b} Figures~\ref{fig:figure02}a and \ref{fig:figure02}c show regular bright-field TEM images, which reveal significant differences at various length scales. While the CB~sample exhibits no distinct structural features on the micrometer scale, alternating regions of bright and dark contrast can be seen for the DCB~sample. Atomic force microscopy~(Figure~S6, Supporting Information) confirms that these alternations are due to the surface topography~(i.e., height variations) rather than the blend composition. Independent of the substrate used, much smoother films with a root-mean-square roughness of $\unit[{\sim}1]{nm}$ were obtained by rapid drying~(CB), as compared to $\unit[{\sim}10]{nm}$ in the case of slow drying~(DCB), in agreement with previous works.\cite{Li2005,Turner2011} 

Instead of that, image contrast on the nanometer scale contains information about the phase separation. Because of the lower density of P3HT~($\unit[{\sim}1.1]{g\,cm^{-3}}$) compared to PCBM~(\unit[1.3]{g\,cm$^{-3}$}), bright areas can be assigned to polymer domains and dark areas to fullerene domains.\cite{Ma2007,Moon2009,vanBavel2009,Dutta2011} The regular TEM~images already suggest a coarser phase separation~(i.e., larger domains) for the CB~cast blend films. However, as these are two-dimensional~(2D) projections of a three-di\-men\-sion\-al~(3D) network, they are  insensitive to possible vertical gradients\cite{vanBavel2009,Campoy-Quiles2008}. To overcome this limitation, we used electron tomography, which yields a volumetric reconstruction of the blend film. Figures~\ref{fig:figure02}b and~\ref{fig:figure02}d show exemplary slices trough the tomograms parallel~($xy$~direction) and perpendicular~($xz$~direction) to the film plane. The tomographic data confirms the trend from the regular TEM imaging. While the CB~sample exhibits well-separated and homogeneously distributed domains in the order of \unit[10]{nm}, the DCB~sample shows a much finer phase separation with a higher degree of interpenetration between the P3HT and PCBM domains. From the $xz$~slices, a columnar structure with transport paths towards the bottom and top of the film can be seen in both cases. Figures~\ref{fig:figure02}e and \ref{fig:figure02}f give an impression of the 3D phase-separated morphology. For this representation, the slices were binarized by attributing each pixel either to the P3HT or PCBM phase~(see also Movie S1 in the Supporting Information). 

\begin{figure}[t]
\includegraphics{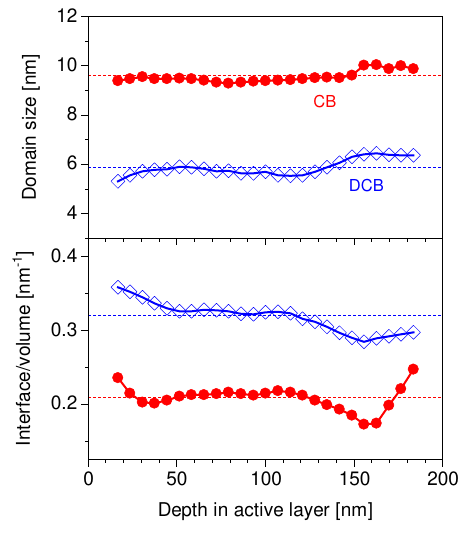}
\caption{Morphological features derived from the electron tomograms. Upper panel: Domain size~$d$ estimated from the width of the self-correlation peak of the radially averaged 2D~autocorrelation function~(see the Supporting Information). Lower panel: Interfacial-area-to-volume ratio of the P3HT phase calculated using Minkowski functionals. Dashed lines indicate the average values.}
\label{fig:figure03}
\end{figure}

For a quantitative statistical analysis of the structural features, we computed the 2D~autocorrelation using the fast Fourier transform~(FFT) algorithm as detailed in the Supporting Information. The upper panel of Figure~\ref{fig:figure03} shows the domain size~$d$, estimated from the width of the autocorrelation function, versus the vertical position. In case of the rapidly dried sample~(CB), the domain size was found to be nearly constant throughout the film, with an average value of $d = \unit[9.6 \pm 0.2]{nm}$. In case of the slowly grown sample~(DCB), the domain size was slightly increasing (from 5.3 to \unit[6.5]{nm}) towards the upper boundary, with an average of $d = \unit[5.9 \pm 0.4]{nm}$. From the binarized slices, we were further able to compute Minkowski functionals such as perimeter and area.\cite{Legland2007} We used these measures to estimate the interfacial-area-to-volume ratio~(Figure~\ref{fig:figure03}, lower panel). As one would expect, the finer phase separation of the DCB~sample is accompanied by a larger amount of interfacial area. We note that due to the 1:1 blend ratio by weight, the P3HT and PCBM phases are not equal in volume. From the number of bright to dark pixels, we estimate the P3HT volume fraction to 0.57~(CB) and 0.54~(DCB), respectively, which fits well to the density ratio mentioned above. Furthermore, the slightly lower P3HT volume for the DCB sample hints to a higher degree of aggregation, as will be discussed in the next section. 

\subsection{Aggregation and Phase Purity}
\label{sec:absorption}
\begin{figure}[t]
\centering
\includegraphics[width=\textwidth]{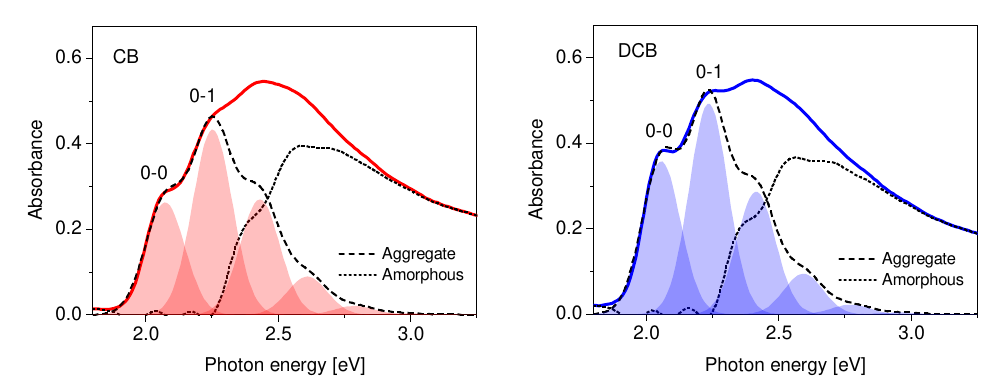}
\caption{Decomposition of the absorption spectra of CB~(left) and DCB~(right) cast blends after thermal annealing. Solid lines are the P3HT component of the P3HT:PCBM absorbance. Dashed lines are fits to the Spano model to the aggregate region~(1.95 to \unit[2.25]{eV}), and shaded areas are the single Gaussian contributions (see Experimental Section for fitting details). Dotted lines represent the residual absorption attributed to amorphous P3HT.}
\label{fig:figure04}
\end{figure}

The casting solvent does not only influence the size of the phase-separated domains, but also the molecular order within them.\cite{Chu2008,Campoy-Quiles2008,Xie2012} To complement our TEM~studies with information about the internal domain structure, we used optical absorption spectroscopy~(Figure \ref{fig:figure04}). Generally, blend films cast from DCB showed more pronounced 0--0 and 0--1~vibronic features, suggesting a higher ordering in the polymer phase.\cite{Brown2003,Tremel2014} It is known that the P3HT~phase in P3HT:PCBM~blend films consists of a mixture of amorphous and aggregated material, the latter formed by lamellar crystallites of 2D~conjugated sheets.\cite{Sirringhaus1999,Moule2008,Tremel2014} For a spectral decomposition of the absorption bands, we fitted the P3HT absorbance component to the model of weakly interacting H-aggregates by Spano and coworkers.\cite{Spano2005,Clark2007,Clark2009} The model treats the absorption in the ordered regions by a series of Gaussian peaks, determined by three parameters: the energy~$E_{0-0}$ of the 0--0 transition, the Gaussian bandwidth~$\sigma$, and the intramolecular free exciton bandwidth~$W$~(see Experimental Section for details).

\begin{table}
\caption{Results of the Spano analysis for CB and DCB cast blend films.}
\begin{tabular}{cccc}
\toprule
Solvent & $E_{0-0}$ (eV) & $\sigma$ (meV) & $W$ (meV)\\
\midrule
CB & 2.048 & 75 & 132 \\
DCB & 2.040 & 73 & 87\\
\bottomrule
\end{tabular}
\label{tab:Spano}
\end{table}

Table~\ref{tab:Spano} and the dashed lines in Figure~\ref{fig:figure04} show the results of the Spano analysis for the spectral region dominated by aggregate absorption. The main difference between the CB and DCB cast blends lies in the exciton bandwidth~$W$, which is known to scale inversely with the intramolecular order in the P3HT aggregates.\cite{Beljonne2000,Gierschner2009} In particular, $W$ has been correlated with the conjugation length, that is, the number of interacting thiophene repeat units. We can estimate the conjugation length using previous approaches,\cite{Gierschner2009,Turner2011,Pingel2010} which gives about 27~repeat units for the CB~blends and 40~repeat units for the DCB~blends. Since both systems were subjected to the same thermal annealing, the apparent differences can only be explained by the solvent evaporation rate. We hypothesize that the fast evaporation of CB gives the P3HT not enough time to organize into larger aggregates, and that thermal annealing cannot fully convert the disordered structure into a highly ordered one with an extended conjugation. This is different to the DCB~blends, where the slow drying already leads to the formation of aggregates of high crystalline quality.\cite{Turner2011} In contrast, $E_{0-0}$ and $\sigma$ are largely unaffected by the solvent. This means that the aggregates in the CB and DCB blends have very similar energetic properties, despite their significantly different expansion in the direction of conjugation. Interestingly, very similar trends with the solvent have recently been reported for a P3HT:NFA blend.\cite{Patel2019} Therefore, we assume that our results are generally valid for P3HT and similar semicrystalline systems.

From the difference between measurement and Spano fit we can calculate also the amorphous absorption component (Figure~\ref{fig:figure04}, dotted lines). Using the ratio between the two absorption fractions, and the extinction coefficients of P3HT in the aggregated and amorphous state,\cite{Clark2009} we estimate the aggregate percentage in the P3HT phase to 45\%~(CB) and 51\%~(DCB), respectively. Even if we cannot resolve the aggregation in our tomography data, there is a direct link to these numbers: Because of the closer packing, aggregated P3HT has a higher density than amorphous P3HT.\cite{Kohn2013} Hence, at a given weight, a slightly smaller P3HT volume is expected in the DCB blends with the higher aggregate percentage. This is exactly the trend we derived from the tomography analysis.

When blended with an acceptor, the P3HT aggregation is directly correlated with the phase purity. In particular, PCBM is known to be intermiscible with P3HT in amorphous state, but not with P3HT in aggregated state.\cite{Treat2011} If we assume that all amorphous P3HT is molecularly mixed with PCBM, we get a refined picture of the morphology consisting of roughly one third pure P3HT, one third pure PCBM and one third mixed phase. For a more accurate estimate, we can combine the P3HT volume from the tomography measurements with the amorphous P3HT fraction from the Spano analysis. This way we arrive at about 31\%~mixed phase in the CB~blends and 26\% in the DCB~blends, which lies in the range that has been reported for annealed P3HT:PCBM blends using high-resolution spectroscopic imaging.\cite{Pfannmoeller2011,Masters2015} Hence, we can conclude that in addition to the higher intramolecular order in the P3HT~aggregates, the DCB~blends also exhibit a higher overall phase purity.

\subsection{Charge Transport}
Having shown that the CB and DCB~blends show clear differences in their morphological features, we now turn to the electrical properties. To estimate the carrier mobilities, we used space-charge limited current~(SCLC) experiments. Figure~S8 in the Supporting Information shows $j$--$V$~curves of electron-only and hole-only devices whose active layers were prepared the same way as for the solar cells. As the simplest and most robust model, we fitted the data in the SCLC~regime to the Mott--Gurney law, and Table~\ref{tab:SCLC} lists the resulting electron and hole mobilities. As the most important result, the magnitude of both $\mu_n$ and $\mu_p$ appears to be fairly independent of the solvent used.

We further analyzed the data in terms of the Gaussian disorder model~(GDM), which explicitly takes into account the hopping nature of transport.\cite{Pasveer2005,Felekidis2018} Also the derived hopping parameters given in Table~\ref{tab:SCLC}, that is, the Gaussian disorder~$\sigma$ and the attempt-to-hop frequency~$\nu_0$, point to very similar transport properties between the CB and DCB~blends. Notably, while there is virtually no difference in the Gaussian disorder for electrons, the disorder for hole transport is $\unit[{\sim}15]{meV}$ smaller in case of the DCB~samples. This shows once again that it is mainly properties of the P3HT phase that are influenced by the solvent. Considering that the apparent transport characteristics are an average over aggregated and amorphous regions, the result is also in line with our hypothesis that the DCB~blends consist of slightly less disordered material. Macroscopically, however, the beneficial effect of the narrower density of states~(lower~$\sigma$) holes are experiencing in the DCB~blends will be canceled out by the slower hopping rate~(lower~$\nu_0$), so that the results from the GDM~model are overall consistent with the Mott--Gurney analysis. This is also in agreement with the general assumption that the macroscopic transport properties are largely dominated by the transport through amorphous regions~(which we assume are not significantly different between the CB and DCB blends) rather than within the aggregates.\cite{Pingel2010,Turner2011,Noriega2013}

\begin{table}[t]
\caption{Charge transport parameters derived from SCLC measurements.}
\begin{tabular}{lcccc}
\toprule
& \multicolumn{2}{c}{Electrons} & \multicolumn{2}{c}{Holes} \\
& CB & DCB & CB & DCB \\
\midrule
Mobility, $\mu$ [$\unit{m^2\,V^{-1}s^{-1}}$] & $1.5 \times 10^{-7}$ & $8.9 \times 10^{-8}$ & $1.6 \times 10^{-8}$ & $1.3 \times 10^{-8}$\\
\addlinespace
Gaussian disorder, $\sigma$ [meV] & 107 & 112 & 79 & 63 \\
Attempt-to-hop frequency, $\nu_0$ [$\unit{s^{-1}}$] & $3 \times 10^{12}$ & $2 \times 10^{12}$ & $9 \times 10^9$ & $7 \times 10^9$\\
\bottomrule
\end{tabular}
\label{tab:SCLC}
\end{table}

It is important to note that SCLC~diodes probe the transport characteristics of charges injected from the contacts. To check whether the results are also relevant for photogenerated charges, we performed resistance-dependent photovoltage~(RPV) measurements on operational solar cells devices. The RPV method is a transient technique with ns to ms resolution, that is, the time scale relevant for charge collection and recombination.\cite{Roland2019} From the RPV~transients shown in Figure~S9~in the Supporting Information, very similar electron mobilities can be derived as from the SCLC~measurements, and about two times lower hole mobilities. As RPV is carried out under much lower carrier densities, the latter might point to a slight carrier density dependence of~$\mu_p$, but is within the typical range when comparing different mobility measurements methods.\cite{Dahlstrom2020} However, the important fact is that there is still no significant difference between the CB and DCB~blends. This shows that also the~(macroscopic) transport of photogenerated charges is only slightly influenced by the differences in morphology. In particular, both the CB and DCB blends show the typical mobility imbalance of about one order of magnitude.\cite{Mihailetchi2006,Bartelt2015} Consequently, the thick devices are supposed to be affected by space charge effects, as will be discussed below.

\subsection{Charge Collection versus Recombination}
\label{sec:tpc}
\begin{figure}
\centering
\includegraphics[width=\textwidth]{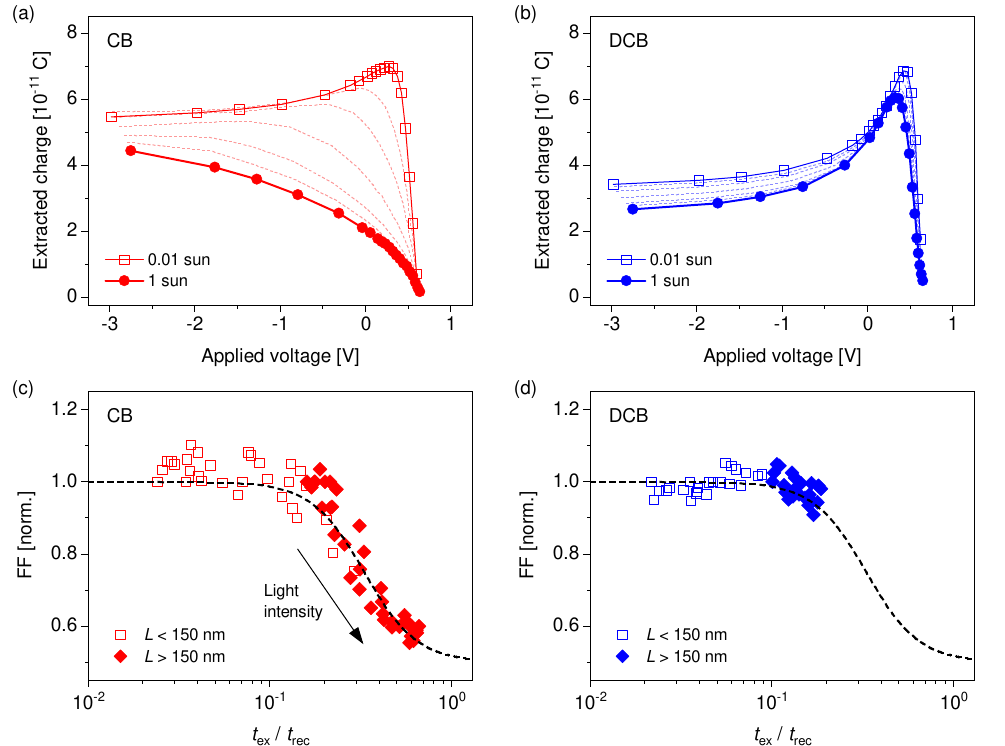}
\caption{Dynamics of charge collection derived from TPC~measurements. (a,b)~Voltage dependence of the extracted charge~$\dq$ for 300-nm thick devices at different background light intensities. (c,d)~Device FF versus the ratio $\tex/\trec$ between the carrier response time in the extraction regime~($V = \unit[-1]{V}$) and the recombination regime~($V \rightarrow \voc$). Data points correspond to 8 samples of different thickness~$L$ and background light intensities ranging from 0.01 to \unit[1]{sun}. The fill factor is normalized to its low-intensity value to exclude other influences such as the quality of the contacts. Dashed lines are a guide to the eye.}
\label{fig:figure05}
\end{figure}

We now focus in more detail on the thickness-dependent competition between charge collection and recombination. To study the dynamics of collection, we measured the transient photocurrent~(TPC) due to a small optical perturbation while the device is held at a constant bias voltage and background illumination. We used a relatively long pulse length of $\unit[100]{\mu s}$ to guarantee that a steady state is reached. Because of the finite carrier mobility, a transient current~$j(t)$ is observed after switching off the light pulse at time $t = 0$, which reflects the carrier sweep-out.\cite{Cowan2011,Li2011} Figure~\ref{fig:figure05}a,b illustrates the TPC~behavior of 300-nm devices by plotting the voltage dependence of the extracted charge,
\begin{equation}
\dq = \int_0^{t_f} j(t)\diffd t,
\end{equation}
where $t_f$ is a time at which charge collection is completed. Data for the whole thickness series can be found in the Supporting Information~(Figures~S10 and S11).

At low background illumination, all devices exhibit a similar behavior, which can be understood as follows. Reducing the internal voltage, $\vint = \voc - V$, by going from reverse to forward bias slows down the current decay and~$\dq$ becomes larger. However, as $V \rightarrow \voc$, the internal field is close to zero and the extracted charge is reduced due to recombination. Altogether, this leads to characteristic maximum of~$\dq$, indicating the point at which recombination starts to compete with collection. Such a voltage dependence proves that the TPC~signals represent the free charge carrier dynamics and are not limited by RC~effects.\cite{Li2011} The situation changes with increasing background illumination. In the thicker CB devices~($L > \unit[150]{nm}$), the extracted charge is drastically reduced and the maximum of~$\dq$ shifts gradually towards higher internal fields. Hence, recombination clearly competes with collection over a large range of voltages. This is in contrast to the DCB~devices, where~$\dq$ is nearly invariant to thickness and light intensity. Also in the thick DCB~devices, recombination is only significant close to~$\voc$, where photogenerated carriers mostly recombine with carriers injected from the contacts.\cite{Wurfel2019}

Figure~\ref{fig:figure05}c,d illustrates the relevance of the TPC~dynamics for the solar cell performance. Shown is the device~FF versus the ratio between the TPC decay time~$\tex$ in the extraction regime~(reverse bias) and $\trec$ in the recombination regime~(close to $\voc$) for a range of thicknesses and light intensities. The ratio~$\tex/\trec$ serves as a figure of merit for the competition between collection and recombination.\cite{Bartesaghi2015,Neher2016} Notably, all data points collapse into a universal curve. For the CB devices, the fill factor drops when~$\tex$ and~$\trec$ are in the same order of magnitude, which is the case for~$L > \unit[150]{nm}$ at high light intensities. In contrast, for the DCB~devices, the absence of data points in this region  indicates that collection is always faster than recombination. Given the very similar mobilities, it is not likely that carriers are collected at a higher rate. Instead, the striking differences between the two systems must be related to the charge recombination. 

\begin{figure}[t]
\centering
\includegraphics{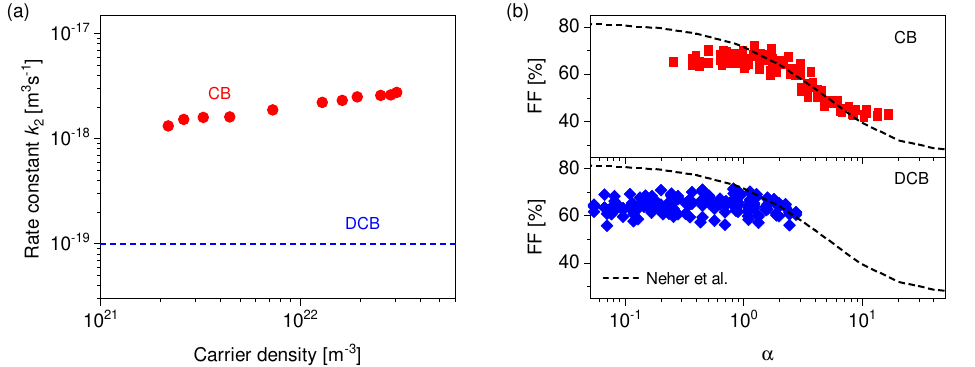}
\caption{Characterization of the recombination kinetics. (a)~Rate constants~$k_2$ for the CB and DCB blends. The data points for the CB~system were taken from a recent literature study\cite{Wilken2020} and confirmed by transient photovoltage measurements~(Supporting Information, Figure~S13). For the DCB system, $k_2$ was estimated as detailed in the text. (b)~Device~FF versus the parameter~$\alpha$ as given in Equation~(\ref{eq:Neher2}) for CB and DCB~solar cells of various thickness and measured at different light intensities. Dashed lines are the expectations according to the modified Shockley model by Neher~et al.\cite{Neher2016}~for recombination rate constants of $k_2 = \unit[2 \times 10^{-18}]{m^3s^{-1}}$~(CB) and $k_2 = \unit[1 \times 10^{-19}]{m^3s^{-1}}$~(DCB), respectively. The degradation of the~FF in the CB~devices can be explained by the larger $k_2$ alone.}
\label{fig:figure06}
\end{figure}

In order to characterize the recombination mechanism, we determined the reaction order~$\delta$ and the ideality factor~$\nid$ using transient and steady-state photovoltage measurements as described in the Supporting Information. For the CB~devices, we find $\delta$ close to~2 and $\nid$ close to~1, which indicates that bimolecular recombination between free electrons and holes is the dominant loss mechanism.\cite{Foertig2012,Kirchartz2012} Hence, we can employ the rate equation~$R = k_2 n^2$ and estimate the rate constant to~$k_2 \approx \unit[2\times10^{-18}]{m^3\,s^{-1}}$~(Supporting Information, Figure~S13). This value confirms recent charge extraction measurements on the same system,\cite{Wilken2020} which also show only a weak dependence on the carrier density~(Figure~\ref{fig:figure06}a).

For the DCB~devices, the apparent recombination behavior is more complex. We find~$\delta$ significantly exceeding~2 and $\nid$ ranging between~1 and~2, which suggests that recombination involves carriers trapped in exponential tails of the density of states.\cite{Kirchartz2011} Given that the DCB~blends actually consist of \textit{less} disordered material than the CB~blends, it does not seem likely that they have more or deeper tail states. Instead, we assume the free carrier recombination~(as given by $k_2$) to be much stronger reduced, so that the trap-assisted regime becomes more apparent. Under these conditions, the transient techniques used herein to determine~$k_2$ do not lead to meaningful results.\cite{Kiermasch2018,Sandberg2019} However, to explain the differences in device performance, we estimate that~$k_2$ must be at least one order of magnitude smaller than in the CB~blends. This is supported by a recent study in which the newly developed impedance-photocurrent device analysis technique showed a rate constant~$k_2$ of about~$\unit[10^{-19}]{m^3\,s^{-1}}$ for comparably processed DCB~devices.\cite{Heiber2018}

To test whether a contrast in $k_2$ alone can explain the striking differences between CB and DCB~devices, we applied the modified Shockley equation by Neher~et~al.,\cite{Neher2016}
\begin{equation}
j = qGL \left\{\exp\left[\frac{q}{(1 + \alpha) \kb T} (V-\voc) \right] - 1\right\},
\label{eq:Neher1}
\end{equation}
where $G$ is the generation rate calculated with our transfer-matrix model and $\alpha$ is a factor that relates charge generation, transport and recombination to each other,
\begin{equation}
\alpha = \frac{q (k_2 G)^{1/2} L^2}{2 \mu_p \kb T}.
\label{eq:Neher2}
\end{equation}
In the denominator we used only the mobility of the slower carrier~(here: holes), which dominates the photocurrent if transport is significantly imbalanced.\cite{Neher2016,Wilken2020} Figure~\ref{fig:figure06}b shows the analytical relationship between the FF and $\alpha$~(dashed lines) together with ca.~200 data points each for the CB and DCB~system, corresponding to different samples of variable thickness~$L$ measured at varying light intensity~(and thus~$G$). There is a reasonable agreement between the Neher model and the experiments, which confirms the validity of the~$k_2$ values. We note that the deviations at low and high $\alpha$~values are due to electrical imperfections~(finite shunt resistance, surface recombination) in the thin devices and space-charge effects in the thick devices, respectively. An important finding from Figure~\ref{fig:figure06}b is that under nearly all  conditions tested, the DCB~devices operate as Shockley-type solar cells~($\alpha < 1$). This is an outstanding result for OPVs, especially with thick active layers.\cite{Armin2018} In contrast, most CB~devices are in the transport-limited regime~($\alpha > 1$).

Figure~S14 in the Supporting Information shows that the Neher model can also correctly describe full $j$--$V$ curves at different light intensities. The notable exception is the thick CB~devices at high intensities. The reason is that under these conditions the photocurrent becomes space-charge limited, which is not considered in Equation~(\ref{eq:Neher1}). It is important to note that due to the similar mobility imbalance, space charge will also build up in the thick DCB~devices, as can be seen from modeled energy band diagrams~(Figure S15, Supporting Information). We have recently shown that collection is limited to the width~$w$ of the space-charge region plus the diffusion length~$L_D$ of the slower charge carrier.\cite{Wilken2020} For the thick devices, we find $w \approx \unit[160]{nm}$ under 1-sun illumination. This gives an alternative way to determine an upper limit for $k_2$ in the DCB blends: The fact that charges are collected from a 300-nm device without noticeable loss means that the hole diffusion length must be at least around~\unit[140]{nm}. This is an exceptionally long diffusion length for~OPVs, outperforming for instance the highly efficient PM6:Y6 system.\cite{Tokmoldin2020} Using the relationship
\begin{equation}
L_D = \sqrt{\frac{\mu_p \tau k_B T}{q}},
\end{equation}
where $\tau = (k_2 G)^{1/2}$, we can then estimate that $k_2$ must be about~$\unit[1 \times 10^{-19}]{m^3\,s^{-1}}$ or less. This is in excellent agreement with our above assumption and confirms that the key difference between the CB and DCB~devices lies in the free charge recombination, which is about 20~times more reduced in the DCB~blends.

\section{Discussion}
\begin{figure}[t]
\centering
\includegraphics[]{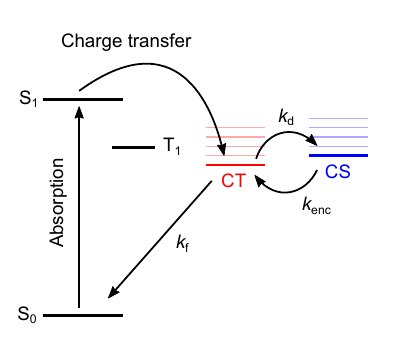}
\caption{Illustration of the relevant energy levels and transitions for charge generation and recombination in organic bulk-heterojunction solar cells. The nongeminate recombination of electrons and holes from the charge separated~(CS) state to the electronic ground state involves the formation of a charge transfer~(CT) pair as intermediate.}
\label{fig:figure07}
\end{figure}

We now want to discuss our findings in the light of recent recombination models and derive design rules for commercially relevant OPV~materials. Figure~\ref{fig:figure07} summarizes the current understanding of charge generation and recombination in organic solar cells.\cite{Shoaee2019,Goehler2018,Burke2015,Murthy2013,Deibel2010b} Both processes involve bound excitons~(either in spin singlet or triplet state), less bound charge transfer~(CT) pairs, and free carriers. Nongeminate charge recombination, on which we will focus in the following, is a two-step process. The first step is the encounter of a free electron and a free hole originating from different photoexcitations~(rate constant~$\kenc$); the resulting  encounter complex has been identified as CT~state with similar properties to the one involved in charge generation.\cite{Murthy2013,Tvingstedt2009} The second step is the decay of the CT~state into the ground state~(rate constant~$k_f$), that is, the actual recombination event. However, instead of decaying into the ground state, the CT~state also has the possibility to dissociate again~(rate constant~$k_d$). Following this rationale, the experimentally observable bimolecular rate constant~$k_2$ can be written as 
\begin{equation}
k_2 = \zetaCT \kenc,
\label{eq:reduction}
\end{equation}
where $\zetaCT$ is a reduction factor related to the CT~kinetics. 

\begin{figure}[t]
\centering
\includegraphics[width=\textwidth]{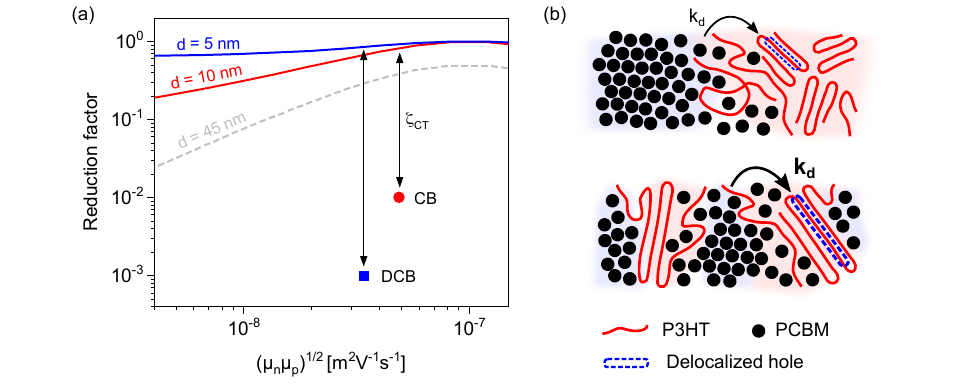}
\caption{Effect of morphology on charge recombination. (a)~Measured reductions~$k_2/k_L$ for the CB and DCB~blends~(data points) in comparison with the model by Heiber~et~al.\cite{Heiber2015} assuming encounter-limited recombination with a reduction factor~$\kenc/k_L$~(lines). Different traces belong to different domain sizes. (b)~Schematic illustration of the morphologies studied herein (top: CB, bottom: DCB). Our data suggests that the more extended delocalization of holes in the DCB blends assists CT states to dissociate.}
\label{fig:figure08}
\end{figure}

When the decay of the CT~state is much faster than its dissociation~($k_f \gg k_d$), the recombination is encounter-limited and $\zetaCT \rightarrow 1$. Hence, in this case, the recombination rate is solely given by~$\kenc$. While in a homogeneous medium, $\kenc$ should be given by the Langevin prefactor~$k_L$, it may become reduced in a blend with phase separation. This is because electrons and holes are confined to different material phases and can only meet at the heterointerface.\cite{Koster2006,Pivrikas2005,Groves2008,Gorenflot2014} Heiber et al.\cite{Heiber2015}~provided a semi-anlytical model for the encounter rate in the presence of phase separation,

\begin{equation}
\kenc = \frac{q}{\varepsilon\varepsilon_0} 2 f_1 \left(\frac{\mu_n^g + \mu_p^g}{2}\right)^{1/g},
\label{eq:Heiber}
\end{equation}
where $f_1$ and $g$ are domain-size dependent factors derived from Monte Carlo simulations on artificial blend morphologies. Figure~\ref{fig:figure08}a illustrates the possible reduction through Equation~(\ref{eq:Heiber}) for a range of mobilities and domains sizes. Comparing the numbers with the measured reduction factors yields two conclusions: First, phase separation cannot explain the differences between the CB and DCB~blends. With the given mobilities and domain sizes, the Heiber model would predict recombination to be (slightly) weaker in the CB~blends, which is the opposite trend to our experimental result. Second, the calculated encounter rates exceed the measured values of~$k_2$ by orders of magnitude, which proves that both the CB and DCB blends are not in the encounter-limited regime. Even in extreme cases, only relatively mild reductions ($>$$10^{-2}$) are predicted, which shows that tuning the domain size is not a promising strategy to strongly suppress charge recombination.

By inserting the calculated encounter rates into Equation~(\ref{eq:reduction}) we can deduce CT~reduction factors of $\zetaCT = 10^{-2}$ and $10^{-3}$ for the CB and DCB blends, respectively. In other words, only 1 in 100 or 1000 encounter events will lead to an actual loss, which implies that the probability for the CT~state to separate again must be much higher than the probability to relax to the ground state~($k_d \gg k_f$). As shown by Burke~et~al.,\cite{Burke2015} this leads to an equilibrium between CT~states and free carriers, reducing the recombination rate by
\begin{equation}
\zetaCT = \frac{k_f}{k_f + k_d}.
\label{eq:Burke}
\end{equation}
Several studies have indicated that the balance between~$k_d$ and~$k_f$ is moved towards separation by the presence of aggregates.\cite{Jamieson2012,Sweetnam2014,Burke2015,Cha2019,Wilken2020b} Energetically speaking, aggregation shifts the molecular orbitals such that the electronic gap is reduced compared to the amorphous state.\cite{Osterbacka2000} In a three-phase morphology, consisting of molecularly mixed, amorphous regions and pure aggregates, this creates an energy cascade pushing carriers from mixed to pure regions. The additional driving force has been shown to not only improve the split-up of CT~pairs during photogeneration, but also to reduce the loss due to charge recombination.\cite{Burke2015,Wilken2020b}

Given the well documented energy level offset between P3HT aggregates and amorphous P3HT mixed with PCBM,\cite{Sweetnam2014} the significant P3HT aggregation in the CB and DCB~blends reasonably explains why both systems are non-Langevin systems. However, the relatively small difference in the P3HT aggregate percentage of~45 to 51\% alone cannot explain why the reduction is one order of magnitude stronger in the DCB~blends. We also find no evidence that the properties of the PCBM are substantially altered by the solvent. Hence, it is reasonable to suggest that the crystalline quality in the P3HT~aggregates, expressed by the exciton bandwidth~$W$, is the decisive factor. Although the parameter~$W$ mainly refers to excitons, a direct connection to the charge domain could be shown.\cite{Clark2009} Here, the significantly smaller value of~$W$ for the DCB blends translates into a larger number of interacting repeating units in the polymer. As a result, the holes are supposed to be more delocalized in the DCB blends. In other words, they have a higher \textit{local} mobility~(as opposed to the macroscopic mobility addressed in the transport measurements), which leads to a higher probability for carriers to overcome the energetic barrier between the interfacial CT~state and the charge-separated state. This in turn increases~$k_d$ and thereby the denominator in Equation~(\ref{eq:Burke}). For a schematic illustration of this scenario, see Figure~\ref{fig:figure08}b.

It should be noted that the CT~dissociation rate will also affect the way how free charges are generated following photoexcitation. Hence, one might argue that the contrast in~$\zetaCT$ is at odds with the similar generation efficiencies we found for the CB and DCB~blends. However, as pointed out by Shoaee~et~al.,\cite{Shoaee2019} equilibrium between CT~states and free charges implies a reverse relationship of the form~$\zetaCT = 1 - \etaCT$, where~$\etaCT$ is the yield of CT~dissociation during charge generation. Varying~$\zetaCT$ from~$10^{-3}$ to~$10^{-2}$, which has drastic consequences for the recombination behavior as shown in this work, corresponds only to a change in~$\etaCT$ from~$0.999$ to~$0.99$. Such small differences in generation efficiency are not distinguishable with the methods used herein. In other words, recombination is much more sensitive to the CT~dissociation rate than generation is.

Another aspect to consider is the role of spin in the recombination. In general, the encounter of two independent charges should form CT states of singlet~($^1$CT) and triplet~($^3$CT) spin state in a 1:3~ratio.\cite{Rao2013,Chow2014,Shoaee2019} The direct transition from $^3$CT to the ground state is spin-forbidden, but the triplet CT~state may undergo back electron transfer to triplet excitons~(T$_1$). Hence, there would be in principle two different decay channels with different relaxation kinetics, which would make the relation between generation and recombination less straightforward.\cite{Shoaee2019} For back electron transfer to be relevant, the T$_1$ level in either the donor or acceptor must be at lower energy than the CT state. Even though such a configuration is not typical for P3HT:PCBM, it cannot be completely ruled out from this work. However, it could be shown that in other materials the loss channel through triplets is turned off upon aggregation.\cite{Rao2013,Chow2014} Thus, if triplets were relevant, it is likely that they would lead to additional losses in the CB rather than in the DCB system, which enforces our view that aggregation is key to suppress charge recombination. Clarification of this aspect would be an interesting direction for future research.

Summarizing, we find that P3HT:PCBM blends processed via the DCB~route display an optimal morphology in terms of reduced recombination and thickness-insensitive device performance. The optimal morphology consists of both amorphous and aggregated regions. To this end, our work confirms earlier suggestions that a three-phase morphology balances best between efficient generation~(mainly to occur in the amorphous phase) and reduced recombination~(carriers are pushed away from the interface towards aggregated regions), and outperforms both purely amorphous and highly ordered blends.\cite{Nyman2015,Schwarz2020} However, the crucial point here is that the mere existence of aggregates in an amorphous matrix is not sufficient to suppress recombination to such an extent that efficient thick-film devices are possible. To achieve this, the aggregates must be of high crystalline quality and purity, which in the present case is only realized in the carefully equilibrated DCB blends. Since the morphological features discussed here are not exceptional but typical for most OPV~blends, we expect our results to be transferable to other systems. In particular, we expect that the guiding principle of having high-quality aggregates is also valid for the acceptor phase, which is supported by a number of recent studies on both fullerenes and NFAs.\cite{Cha2019,Wilken2020b,Liu2020,Hosseini2020}

\section{Conclusions}
In this paper, we revisited the classical P3HT:PCBM blend to establish connections between the nanoscale morphology and device physics of organic bulk-heterojunction solar cells. By exploiting a structure--property relationship that has so far received little attention, we could show that aggregation is the key feature to reduce recombination losses. However, in order to reduce the recombination rate to such an extent that the solar cells operate as Shockley-type devices even in thick junctions, the mere presence of aggregates is not sufficient. For this to be the case, the aggregates must be of high crystalline quality and purity, so that the charge carriers are delocalized over larger areas. The delocalization boosts the dissociation of charger-transfer states that are formed by the encounter of free electrons and holes. In the case of P3HT:PCBM, such a situation is realized in carefully equilibrated blend films that are slowly dried after spin-coating. The optimized blends show extraordinarily long hole diffusions lengths exceeding~\unit[100]{nm} and can also tolerate the build-up of space charge due to imbalanced transport.

In contrast to the order and energy landscape within the phases, we find that the structure size of the phase separation plays only a minor role in charge recombination. The fact that charge carriers are confined in donor and acceptor domains, so that a coarser phase separation would be preferable from a geometric point of view, is far outperformed by the effect of aggregation. Therefore, optimization of the domain size, for example through nanostructuring, is not a promising approach to significantly suppress recombination. Instead, the focus should lie on molecular order, crystalline quality and phase purity. This is especially relevant for devices based on nonfullerene acceptors, where a reduction of the recombination rate is crucial due to the typically low carrier mobilities.

\section{Experimental Section}
\paragraph{Materials} Regioregular P3HT was purchased from Rieke Metals (4002-E, molecular weight 50--70 kDa, regioregularity 91--94\%). PCBM was purchased from Solenne BV (purity 99.5\%). Chlorobenzene (CB), 1,2-dichlorobenzene (DCB) and poly\-ethylen\-imine (PEIE) were purchased from Sigma-Aldrich. Poly(3,4\-ethylene\-di\-oxy\-thio\-phene) poly\-styrene sulfonate (PEDOT:PSS) was purchased from Heraeus (Clevios P VP AI 4083). Indium tin oxide~(ITO) covered glass substrates were purchased from Pr\"{a}zisions Glas \& Optik.

\paragraph{Blend-Film Preparation} Blend solutions were prepared by dissolving P3HT and PCBM in 1:1 weight ratio in either CB or DCB and stirred at \unit[60]{$^\circ$C} for \unit[12]{h} prior to further processing. Blend films were produced by dynamic spin coating. The thickness was controlled by the concentration of the solution~(30 to \unit[60]{mg/ml}) and the spin-coating speed (500 to \unit[1500]{rpm}). After complete drying in a closed vessel, all samples were thermally annealed at \unit[150]{$^\circ$C} for \unit[10]{min}. All preparation was carried out under a dry nitrogen atmosphere. 

\paragraph{Electron Tomography} Specimens for TEM were prepared by depositing blend films on glass substrates covered with a sacrificial layer of~PEDOT:PSS. After immersion in deionized water, the PEDOT:PSS layer dissolved, and the free-floating blends were transferred to 300-mesh copper grids. Bright-field TEM images were acquired with a 200-kV field emission electron microscope~(Jeol JEM-2100F) at an underfocus of \unit[10]{$\mu$m}~\cite{Ma2007,Moon2009,Dutta2011}. Electron tomography was performed by recording a series of TEM~images under different viewing angles by tilting the specimen in a range of~$\pm65^\circ$ in nonequidistant increments according to Saxton~\textit{et~al.}~\cite{Saxton1984}. The tilt series was assembled to a volumetric reconstruction~(voxel size \unit[0.43]{nm}) using the simultaneous iterative reconstruction technique~(SIRT) with 25~iterations. Acquisition, alignment, and reconstruction of tomographic data was done with the software package Temography~(System in Frontier, Inc.). For the FFT analysis, the reconstructed volumes were cut into series of horizontal slices. Binarization of the grayscale slices was done by applying a median filter~($\unit[9 \times 9]{pixels}$) and thresholding using Otsu's method.

\paragraph{Device Fabrication} Inverted solar cells were fabricated by spin coating a 50-nm~layer of ZnO nanoparticles~(diameter \unit[5]{nm}, see Ref.~\citenum{Wilken2014} for details) on cleaned and patterned ITO~substrates. Subsequently, the active layer was deposited either from CB or DCB as described above. After thermal annealing, a \ce{MoO3}~(\unit[12]{nm})\slash{}\ce{Ag}~(\unit[150]{nm}) electrode was evaporated under high vacuum~(\unit[$10^{-6}$]{mbar}). The active area was \unit[0.3]{cm$^2$}. Solar cells were encapsulated with glass slides and an UV-cured optical adhesive. Single-carrier diodes were fabricated with the device architecture ITO\slash{}PEIE\slash{}P3HT:PCBM\slash{}Ca\slash{}Al (electrons) and ITO\slash{}PEDOT:PSS\slash{}P3HT:PCBM\slash{}\ce{MoO3}\slash{}Ag (holes), respectively.

\paragraph{Characterizations} Current--voltage curves were recorded with a parameter analyzer~(Keithley 4200). A class AAA solar simulator (Photo Emission Tech) was used to provide simulated AM1.5G~illumination at~\unit[100]{mW cm$^{-2}$}. The EQE was measured with a custom-built setup~(Bentham PVE300), equipped with a 75-W~Xe arc lamp and a monochromator. Photocurrent signals were modulated at~\unit[780]{Hz} and monitored with a lock-in amplifier~(Stanford Research Systems SR830). White-light bias illumination was provided by a 50-W~halogen lamp. The intensity of all light sources used was calibrated with a KG5-filtered reference solar cell. UV--vis absorption spectra were recorded from optically thin films on glass with a spectrophotometer~(Varian Cary 100) and corrected for the transmission of the substrate. The P3HT absorption component was fitted to the Spano model,\cite{Spano2005,Clark2007,Clark2009} which treats the absorption spectrum as a series of Gaussian bands,
\begin{equation}
A \propto \sum_{m = 0} \left(\frac{S^m}{m!}\right) \left(1 - \frac{W \text{e}^{-S}}{2 E_p} \sum_{n \neq m} \frac{S^n}{n! (n-m)}\right)^2 \exp \left( - \frac{\left(\hbar\omega - E_{0-0} - mE_p - \frac{1}{2}\frac{W S^m \text{e}^{-S}}{m!}\right)^2}{2\sigma^2} \right),
\label{eq:Spano}
\end{equation}
where $\hbar\omega$ is the photon energy, $S$ the Huang-Rhys factor, $E_p$ the intramolecular vibrational energy, $E_{0-0}$ the energy of the 0--0 transition, $\sigma$ the Gaussian bandwidth, and $W$ the intramolecular free exciton bandwidth. Assuming $S = 1$ and $E_p = \unit[0.179]{eV}$ for the C=C symmetric stretch,\cite{Clark2007} the only free fit parameters were $E_{0-0}$, $\sigma$ and $W$. Film thicknesses were measured with a stylus profiler~(Veeco Dektak 6M).

\paragraph{Transient Measurements} For TPC and TPV~measurements, a 4-W~white-light LED (Seoul P4) was used to provide constant background illumination. Another LED~(wavelength \unit[525]{nm}, \unit[250]{ns} rise/fall time), driven by a double pulse generator~(Agilent 81150A), was used to apply a small optical perturbation to the sample. The second channel of the pulse generator served as bias-voltage source. Current and voltage transients were recorded with a 1-GHz~digital storage oscilloscope~(Tektronix DPO7104) at an input impedance of~\unit[50]{$\Omega$} and~\unit[1]{M$\Omega$}, respectively. The TPC~transients were routinely corrected for RC time effects as described elsewhere.\cite{Kettlitz2013} A biased silicon detector~(Thorlabs DET36A) was used to monitor the switching dynamics and background light intensity. The latter was pre-adjusted for each device by matching the current--voltage response under simulated sunlight. Light sources were attenuated with neutral density filters. Experiments were done at room temperature and ambient pressure.

\paragraph{Transfer-Matrix Model} One-dimensional transfer-matrix calculations were performed with a customized MATLAB code.\cite{Burkhard2010} The optical constants of all materials involved were determined by spectroscopic ellipsometry. The validity of the model was checked by comparing simulated and experimental reflectance spectra of complete OPV devices. Further details are given in the Supporting Information.

\begin{acknowledgement}
We thank Michael Koopmeiners for preliminary experiments and AFM~measurements, and Edit Kieselhorst, Erhard Rhiel, Matthias Macke and Dirk Otteken for technical support. We are also thankful to Carsten Deibel, Oskar J.~Sandberg, Manuela Schiek, Janet Neerken, Konstantin Kloppstech, Nils K\"onne, Nora Wilson, Christian Ahl\"ang and Christoph Norrenbrock for helpful discussions during various stages of this work. S.W. acknowledges funding through the Research Mobility Programme of \AA{}bo Akademi University. D.S. is grateful for financial support from the Magnus Ehrnrooth Foundation. S.D. acknowledges funding from the Society of Swedish Literature in Finland. M.N. acknowledges funding from the Jane and Aatos Erkko Foundation through the ASPIRE project. This project has received funding from the European Union's Horizon 2020 research and innovation programme under the Marie Sk\l{}odowska-Curie grant agreement No~799801~(``ReMorphOPV'').
\end{acknowledgement}

\bibliography{ms}

\end{document}

\clearpage

\tableofcontents

\clearpage

\section{Summary of Literature Studies}
\begin{table}
\small
\begin{flushleft}
\caption{Recombination studies on annealed P3HT:PCBM solar cells.}
\label{tab:tableS1}
\begin{tabular}{lllcl}
\toprule
Study &  Solvent & Annealing & Reduction factor~$\zeta$ & Experiment\\
\midrule
%\multicolumn{4}{l}{\textbf{Langevin reduction factor}}\\
Pivrikas~et~al.~\cite{Pivrikas2005} (2005) & & & $\sim10^{-4}$ & TOF \\
Ferguson~et~al.~\cite{Ferguson2011} (2011) & CF & solvent-vapor & $3 \times 10^{-4}$ & TRMC\\
Ju\v{s}ka~et~al.~\cite{Juska2005} (2005) & & &  $5\times10^{-4}$ & DI\\
Bartelt~et~al.~\cite{Bartelt2015} (2015) & CF & thermal &  $7\times10^{-4}$ & IV\\
Deibel~et~al.~\cite{Deibel2008} (2008) & CB & thermal & $\sim10^{-3}$ & CELIV \\
Kniepert~et~al.~\cite{Kniepert2014} (2014) & CF & thermal & $1\times10^{-3}$ & BACE \\
Heiber~et~al.~\cite{Heiber2018} (2018) & DCB & solvent-vapor & $1\times10^{-3}$ & IPDA \\
Shuttle~et~al.~\cite{Shuttle2010} (2010) & Xylene & thermal & $2\times10^{-3}$ & CE\\
Shoaee~et~al.~\cite{Shoaee2019} (2019) & DCB & solvent-vapor & $2\times10^{-3}$ & BACE\\
Wetzelaer~et~al.~\cite{Wetzelaer2013} (2013) & CF & thermal & $3\times10^{-3}$ & IV\\
Kniepert~et~al.~\cite{Kniepert2011} (2011) & DCB & solvent-vapor & $6\times10^{-3}$ & TDCF\\
Shuttle~et~al.~\cite{Shuttle2008c} (2008) & CB & thermal & $10^{-2}$--$10^{-3}$ & TA\\
Garcia-B.~et~al.~\cite{Garcia-Belmonte2010} (2010) & DCB & & $\sim10^{-2}$ & EIS \\
Guo~et~al.~\cite{Guo2010} (2010) & CB & thermal & $1\times10^{-2}$ & TA\\
Mingebach~et~al.~\cite{Mingebach2012} (2012) & CB & thermal & $2\times10^{-2}$ & TDCF\\
Mauer~et~al.~\cite{Mauer2011} (2011) & CB & thermal & $10^{-1}$--$10^{-3}$ & DP\\
\bottomrule
\end{tabular}

\begin{tabular}{lll}
& CB & Chlorobenzene\\
& CF & Chloroform\\
& DCB & 1,2-Dichlorobenzene\\
\addlinespace
& BACE & Bias-assisted charge extraction\\
& CE & Charge extraction\\
& CELIV & Charge extraction by linearly increasing voltage\\
& DI & Double injection\\
& DP & Double-pulse technique with variable delay\\
& EIS & Electrical impedance spectroscopy\\
& IPDA & Impedance-photocurrent device analysis\\
& IV & Current-voltage analysis\\
& TA & Transient absorption\\
& TDCF & Time-delayed collection field\\
& TOF & Time of flight\\
& TRMC & Time-resolved microwave conductivity\\
\end{tabular}
\end{flushleft}
\end{table}

\clearpage

\begin{figure}
\centering
\includegraphics[width=0.8\textwidth]{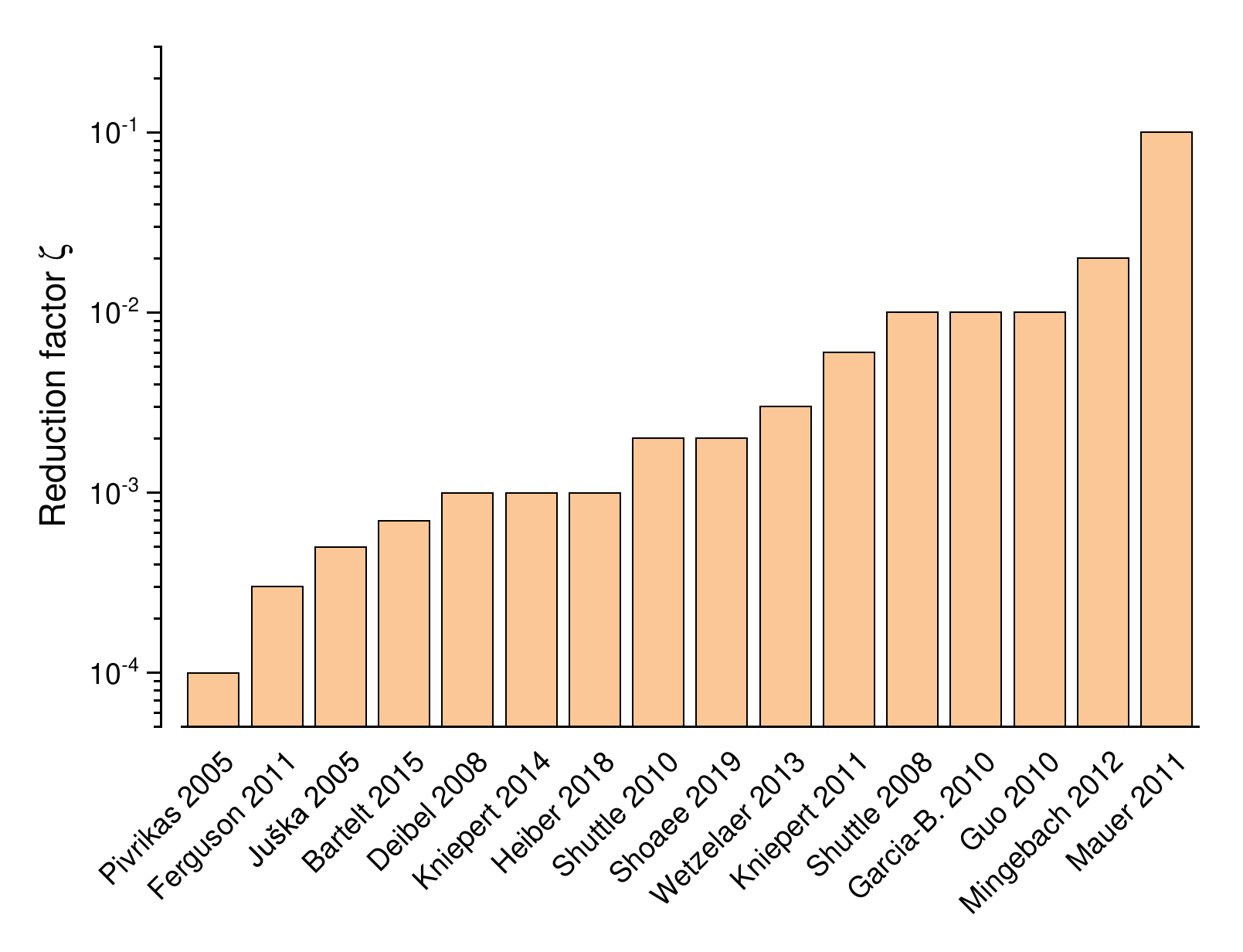}
\caption{Reported values of the reduction factor~$\zeta$ for annealed P3HT:PCBM solar cells. See Table~\ref{tab:tableS1} for details on the literature studies. For studies that specify a range, the upper bound is given in the diagram.}
\label{fig:figureS1}
\end{figure}

\clearpage

\section{Solar Cell Device Performance}

\begin{table}
\caption{Photovoltaic performance of the CB and DCB thickness series.$^\dagger$ }
\begin{tabular}{cccccc}
\toprule
Solvent & Thickness & $\voc$ & $\jsc$ & FF & PCE \\
 & (nm) & (\unit{mV}) & ($\unit{mA/cm^2}$) & & (\%) \\
 \midrule
CB & 65	& 635	& 7.5	& 0.62 & 3.0 \\
 & 100	& 621 & 6.9 &	0.64 & 2.7 \\
 & 140	& 618	& 7.4 &	0.60 & 2.7 \\
 & 150	& 616	& 8.1 &	0.50 & 2.5 \\
 & 190	& 606	& 8.6 & 0.44 & 2.3 \\
 & 220	& 601	& 7.8 & 0.43 & 2.0 \\
 & 260	& 595	& 6.0 & 0.42 & 1.5 \\
 & 300	& 589	& 6.0 & 0.42 & 1.5 \\
 & 350	& 588	& 4.5 & 0.43 & 1.1 \\
\addlinespace
DCB & 45 & 646 & 7.4 & 0.60 & 2.9 \\
 & 70 & 640 & 8.2 & 0.61 & 3.3 \\
 & 105 & 616 & 7.4 & 0.56 & 2.7 \\
 & 140 & 630 & 7.6 & 0.58 & 2.8 \\
 & 180 & 629 & 8.9 & 0.61 & 3.4 \\
 & 200 & 622 & 9.3 & 0.58 & 3.4 \\
 & 250 & 625 & 8.8 & 0.60 & 3.3 \\
 & 300 & 624 & 9.1 & 0.60 & 3.4 \\
 & 320 & 620 & 9.3 & 0.60 & 3.5 \\
 \bottomrule
\end{tabular}

$^\dagger$Data was obtained under simulated AM1.5G solar irradiation of $\unit[100]{mW/cm^2}$ with a spectral mismatch factor of 1.015--1.032~(depending on sample). The reported values for each thickness represent an average over 4~devices.
\label{tab:IVparams}
\end{table}

\clearpage
\section{Transfer Matrix Model}
\begin{figure}[h]
\centering
\includegraphics[width=\linewidth]{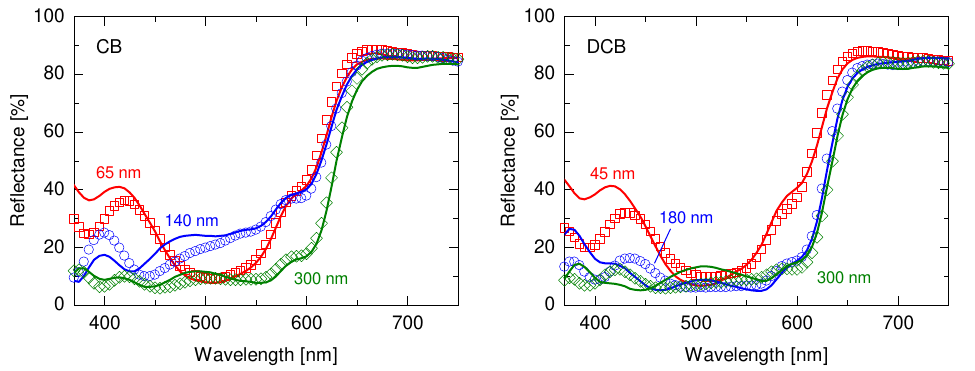}
\caption{Comparison of the experimental~(data points) and simulated~(solid lines) reflectance of complete solar cell devices with variable active-layer thickness.}
\label{fig:R_exp_vs_sim}
\end{figure}

Optical characteristics of the solar cells were modeled using a one-dimensional transfer-matrix approach.\cite{Pettersson1999,Burkhard2010} Each material layer involved in the device stack was treated as homogeneous medium, characterized by its film thickness and complex index of refraction~$\tilde{n} = n' + \text{i}n''$, where $n'(\lambda)$ and $n''(\lambda)$ are the refractive index and extinction coefficient, respectively. The latter were experimentally determined by means of variable angle spectroscopic ellipsometry. Therefore, the ellipsometric parameters $\Psi(\lambda)$ and $\Delta(\lambda)$ were recorded under different angles of incidence with a rotating-analyzer instrument~(Woollam VASE) and fitted to adequate dispersion models. Further details regarding the analysis can be found in previous publications.\cite{Scheunemann2015,Wilken2015} Figure~\ref{fig:R_exp_vs_sim} illustrates the quality of the optical model by comparing measured and simulated reflection spectra of complete solar cell devices.

\clearpage
\section{Internal Quantum Efficiency}

\begin{figure}
\centering
\includegraphics[width=\textwidth]{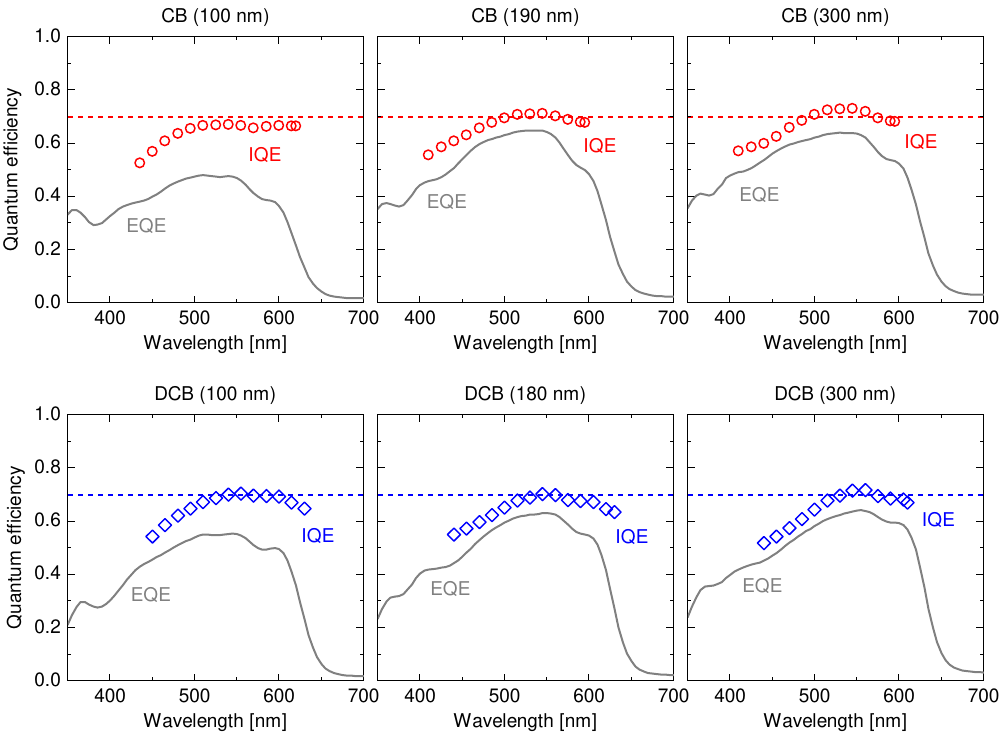}
\caption{Internal quantum efficiency~(IQE) for devices of different thickness as calculated from the measured EQE without white-light bias, the measured reflectance~($R$) and the modeled parasitic absorption~(PA) using the relation $\text{IQE} = \text{EQE}/(1 - R - \text{PA})$. The gradients seen in the IQE at low wavelengths are due to incomplete exciton harvesting in the PCBM phase.\cite{Burkhard2009} Horizontal dashed line indicate an IQE of 0.7 as it was assumed in the photocurrent simulations in the main text.}
\label{fig:figureS1}
\end{figure}

\clearpage
\section{Light-Intensity Dependent Measurements}

\begin{figure}[h]
\centering
\includegraphics[width=\textwidth]{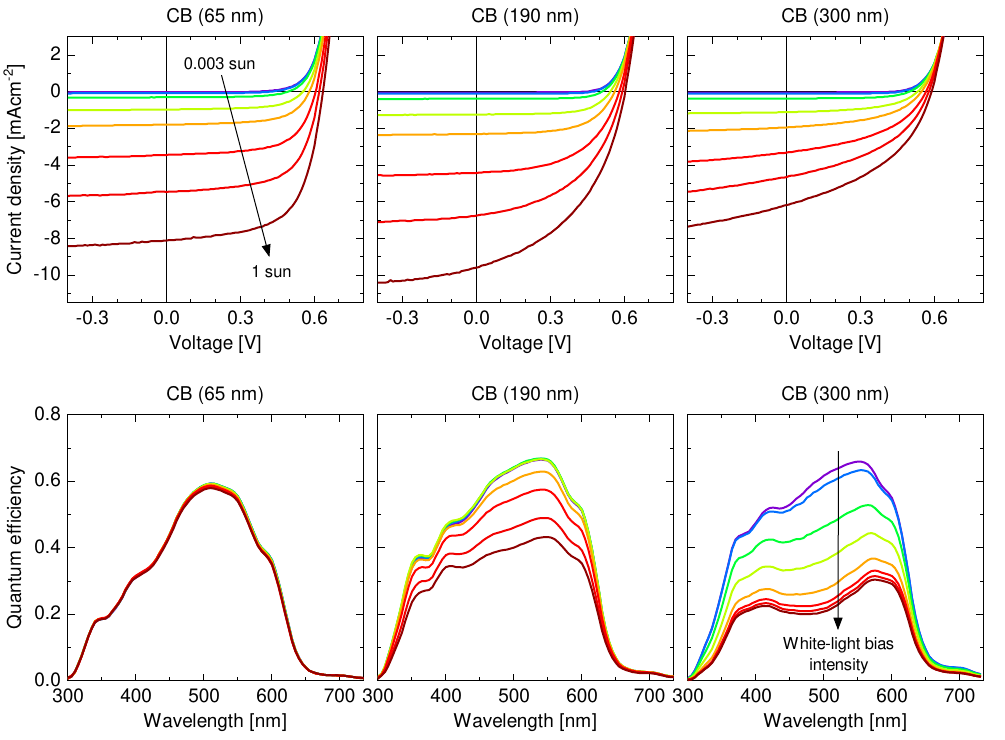}
\caption{Light-intensity dependent $j$--$V$ curves~(upper row) and corresponding white-light biased EQE spectra~(lower row) for CB devices of variable active-layer thickness as indicated in the figure. Arrows indicate increasing~(bias) light intensity.}
\label{fig:IV_EQE_CB}
\end{figure}

\begin{figure}[h]
\centering
\includegraphics[width=\textwidth]{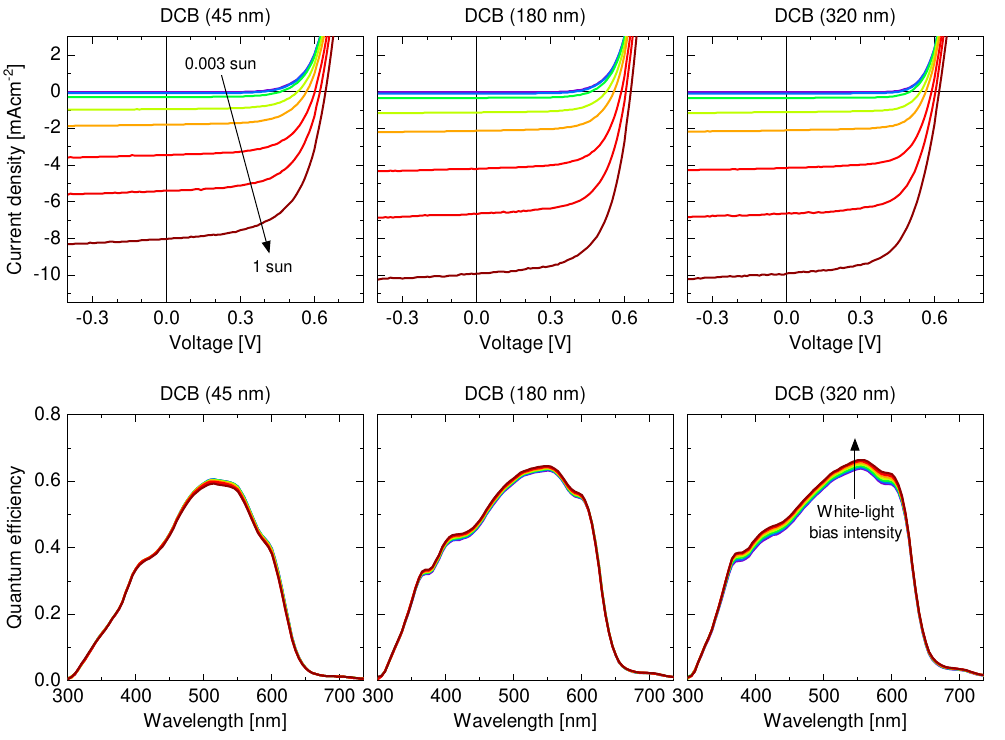}
\caption{Same representation as in Fig.~\ref{fig:IV_EQE_CB} for DCB devices of variable active-layer thickness. The slight improvement of the EQE with increasing light intensity seen in the thicker devices may be a consequence of trap-filling.}
\label{fig:IV_EQE_DCB}
\end{figure}

\clearpage
\section{Atomic Force Microscopy}

\begin{figure}[h]
\centering
\includegraphics[width=\textwidth]{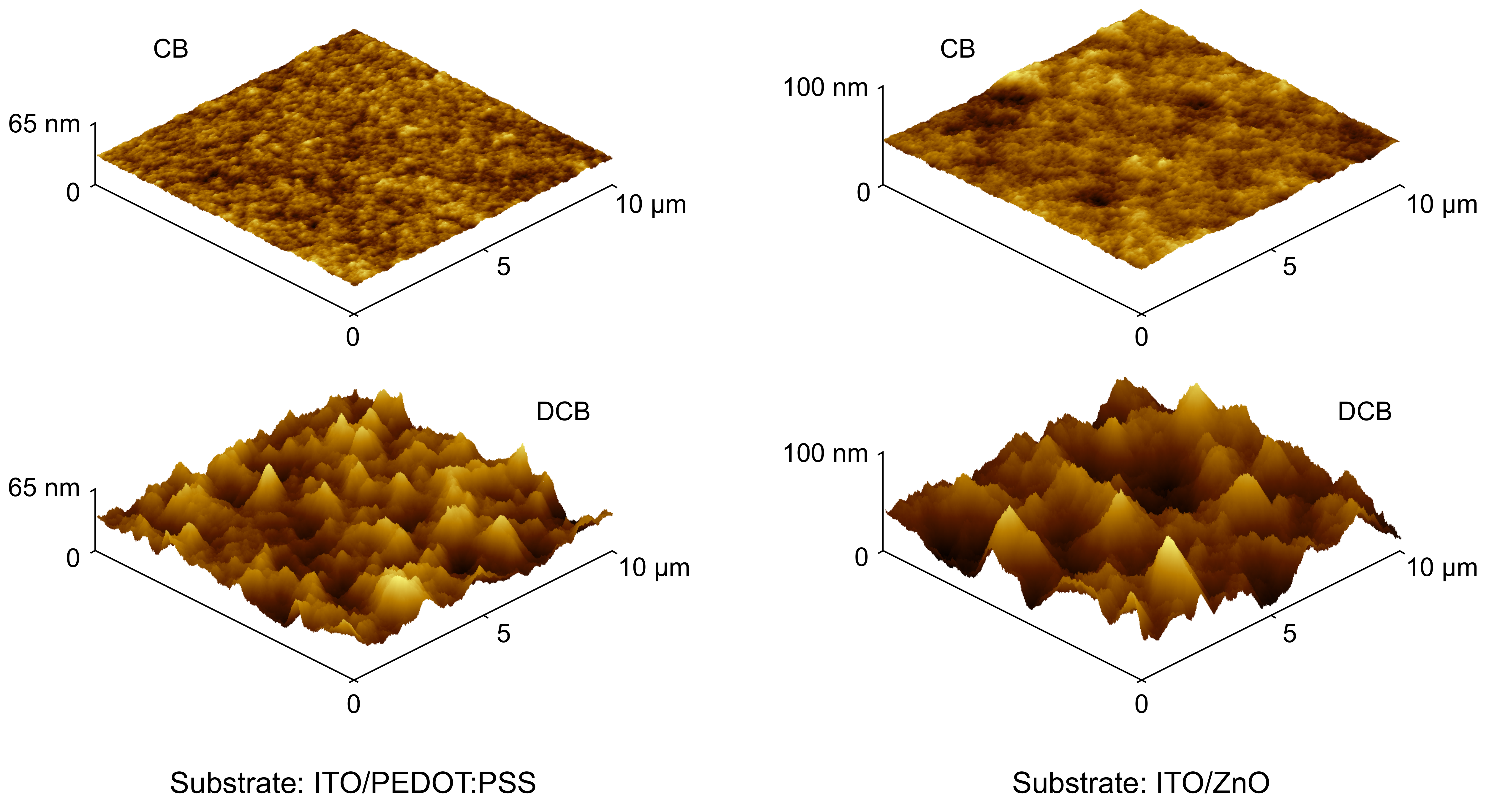}
\caption{Surface topography obtained from atomic force microscopy on CB and DCB blend films~(thickness: \unit[250]{nm}) coated on different types of substrates. Independent of the substrate used, processing from DCB results in much rougher films on the $\unit{\mu m}$~scale. The values of the root-mean-square roughness calculated from the $\unit[10 \times 10]{\mu m}$ scans are \unit[0.9]{nm}~(CB) and \unit[9.2]{nm}~(DCB) for the samples on ITO/PEDOT:PSS, and \unit[1.5]{nm}~(CB) and \unit[14]{nm}~(DCB) for the samples on ITO/ZnO.}
\label{fig:AFM}
\end{figure}

\clearpage
\section{Fourier-Transform Analysis}
For a quantitative statistical analysis of the morphological features revealed by electron tomography, we calculated the 2D autocorrelation function~(ACF) for each slice parallel to the film plane. For an $M \times N$ image, the ACF is given by
\begin{equation}
\Psi_{gg}(m,n) = \sum_{k=1}^{M-m}\sum_{\ell=1}^{N-n} g_{m,n}\,g_{m+k,n+\ell},
\label{eq:ACF}
\end{equation}
where $g_{m,n}$ denotes the brightness value at pixel position~$(m,n)$. The sums in Equation~(\ref{eq:ACF}) were evaluated in the reciprocal space using the fast Fourier transform algorithm. Assuming that the phase-separated domains are randomly oriented in the lateral plane, it is useful to introduce a radially averaged ACF by transformation to polar coordinates~$(r,\varphi)$ and integration over the polar angle,
\begin{equation}
\left\langle\Psi_{gg}\right\rangle_\varphi = \int_0^{2\pi} \Psi_{gg}(r\cos\varphi,r\sin\varphi)\diffd\varphi.
\label{eq:rACF}
\end{equation}

\begin{figure}
\centering
\includegraphics[width=\textwidth]{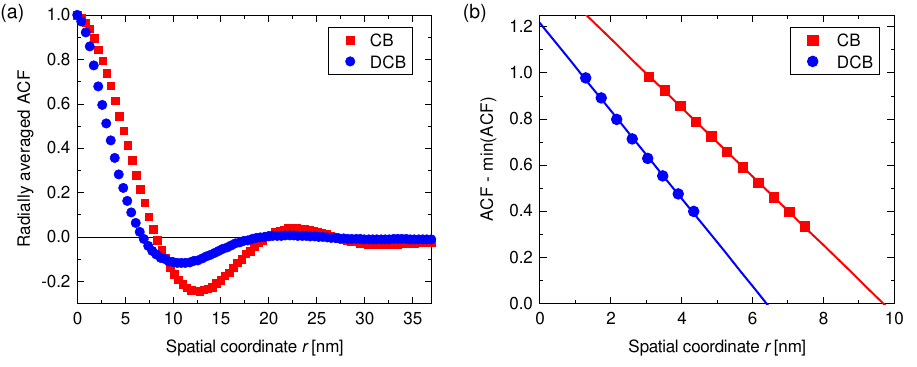}
\caption{Statistical analysis of the tomography data. (a)~Radially averaged ACF for an exemplary $xy$~slice through the reconstructed volumes. (b)~Linear part of the self-correlation peak with the ordinate shifted by the first minimum value of the ACF. The domain size was derived from the zero intercept of a linear fit to the data~(straight lines).}
\label{fig:ACF}
\end{figure}

Figure~\ref{fig:ACF}a shows the results of Equation~(\ref{eq:rACF}) for an exemplary slice from the middle of the tomograms. The data is normalized to~$-1,1$, where 1 indicates perfect correlation and $-1$ perfect anti-correlation. The shape of the ACFs is typical of a periodic two-phase system  distorted by domain size fluctuations, long-spacing variations, and diffusive phase boundaries.\cite{Strobl1980} While the first side maximum is related to the period of the pseudo-periodic structure, the width of the self-correlation peak centered at~$r = 0$ contains information about the average domain size~$d$. According to Strobl and Schneider,\cite{Strobl1980} the domain size was estimated by extrapolating the linear region to the baseline intercept~(Figure~\ref{fig:ACF}b).

\clearpage
\section{Charge Transport Measurements}
\begin{figure}
\centering
\includegraphics[width=\textwidth]{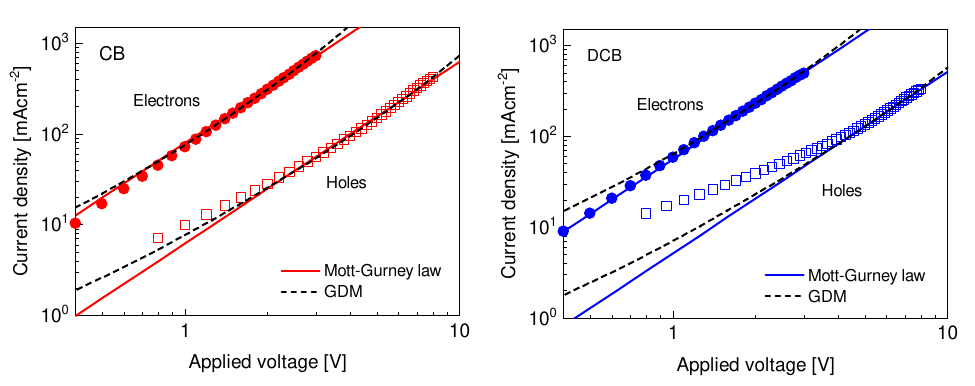}
\caption{Charge transport in CB~(left) and DCB~(right) cast blends. Shown are typical SCLC~current--voltage characteristics of electron-only~(closed symbols) and hole-only~(open symbols) devices. Straight lines are fits to the Mott--Gurney law, $j = 9\varepsilon\varepsilon_0 \mu_{n,p} V^2/8L^3$. Dashed lines are the result of drift--diffusion simulations using the extended Gaussian disorder model~(GDM, see Ref.~\citenum{Felekidis2018} for details).}
\label{fig:figure05}
\end{figure}

\begin{figure}
\centering
\includegraphics[width=\textwidth]{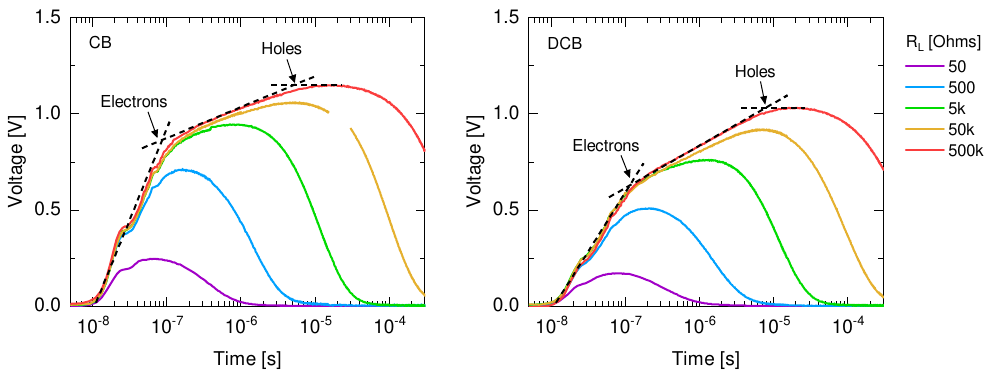}
\caption{Resistance-dependent photovoltage~(RPV) transients for complete solar cell devices at about \unit[180]{nm} thickness. The RPV~experiments were carried out with a pulsed Nd:YAG laser~(wavelength \unit[532]{nm}) and a variable load resistance~$R_L$ as indicated in the figure. From the photovoltage shoulders, which corresponds to the transit times of electrons and holes, respectively, very similar carrier mobilities of $\mu_n \approx \unit[10^{-3}]{cm^2\,V^{-1}s^{-1}}$ and $\mu_p \approx \unit[5 \times 10^{-5}]{cm^2\,V^{-1}s^{-1}}$ can be estimated for the CB and DCB sample. While the electron mobility agrees well with the SCLC experiments, the hole mobility is about two times smaller. This might indicate a slight carrier density dependence of the hole transport as RPV is performed under much lower carrier densities than SCLC, but is within the typical range when comparing different mobility measurement methods.\cite{Dahlstrom2020}}
\label{fig:figureS1}
\end{figure}

\clearpage
\section{Transient Photocurrent}

\begin{figure}
\centering
\includegraphics[width=0.95\textwidth]{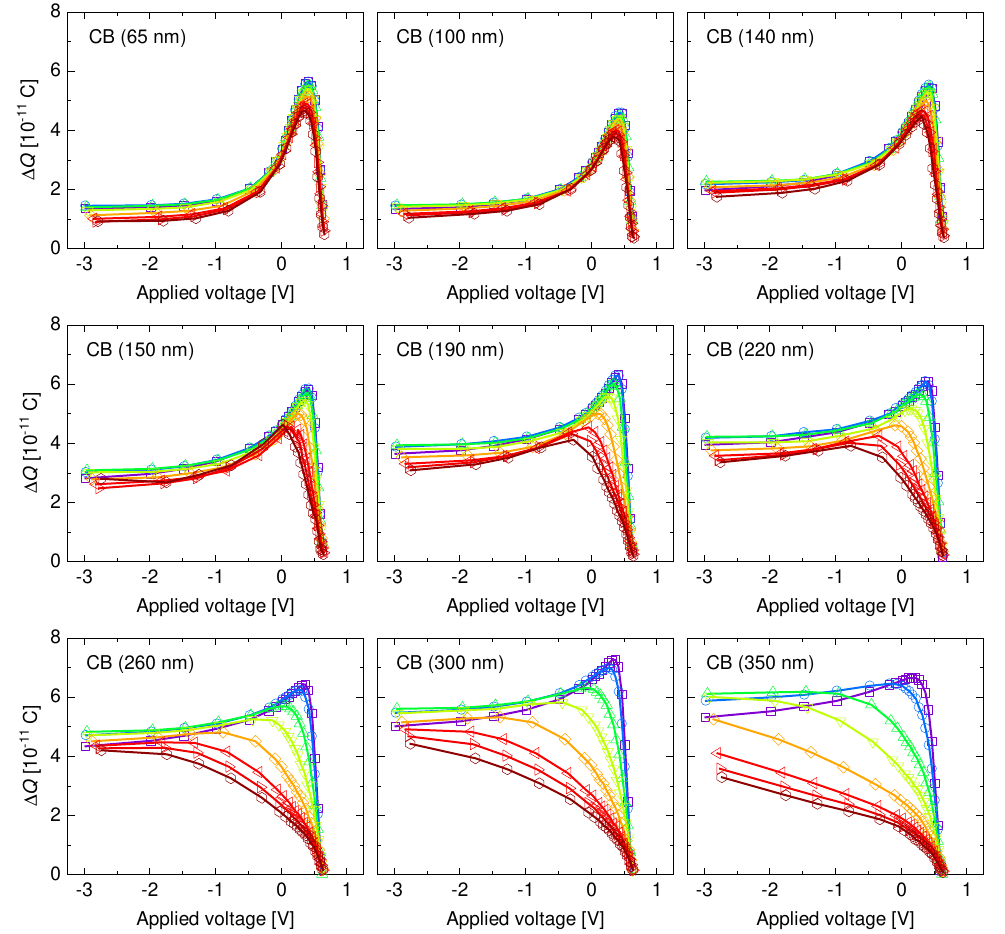}
\caption{Extracted charge~$\dq$ versus applied voltage for CB devices of variable thickness, ranging from \unit[65]{nm}~(top left corner) to \unit[350]{nm}~(bottom right corner), in dependence of the background illumination intensity. The voltage axis was corrected for the voltage drop due to the current offset generated by the background light.}
\end{figure}

\begin{figure}
\centering
\includegraphics[width=0.95\textwidth]{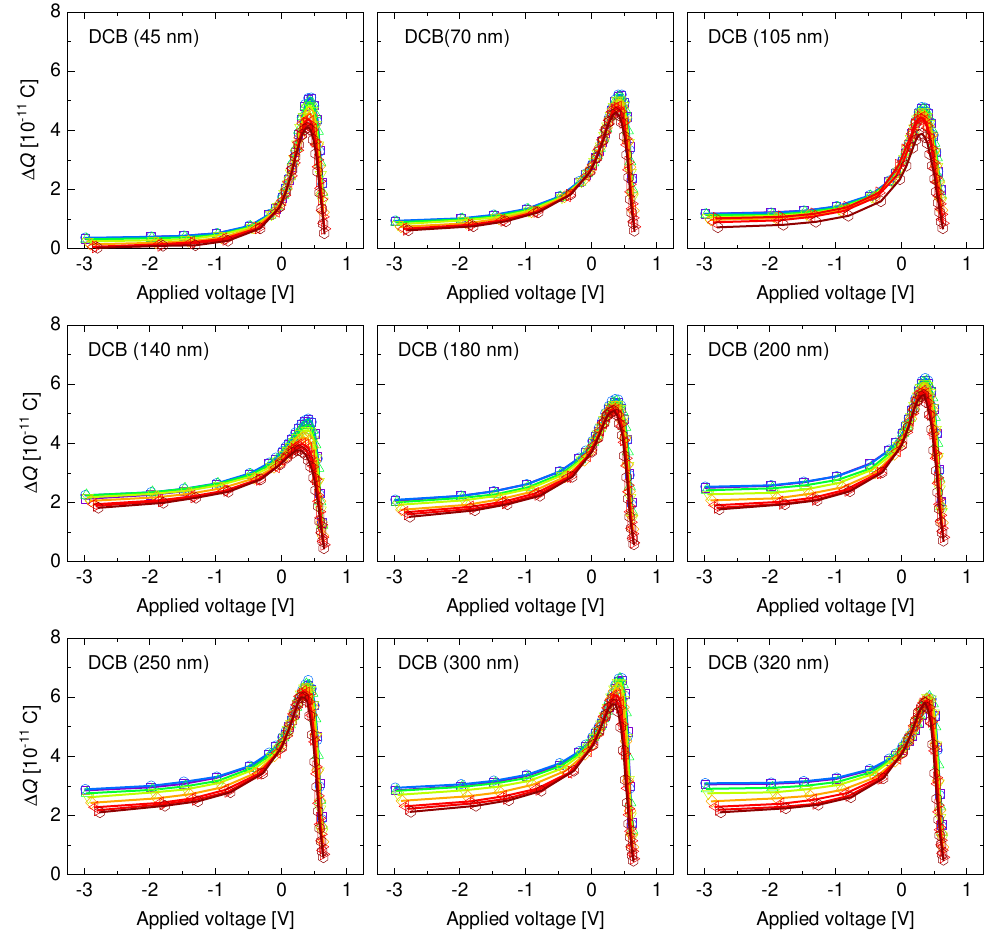}
\caption{Extracted charge~$\dq$ versus applied voltage for DCB devices of variable thickness, ranging from \unit[45]{nm}~(top left corner) to \unit[320]{nm}~(bottom right corner), in dependence of the background illumination intensity. The voltage axis was corrected for the voltage drop due to the current offset generated by the background light.}
\end{figure}

\clearpage
\section{Reaction Order and Light Ideality Factor}
Two parameters that serve as fingerprints for the recombination mechanism are the reaction order~$\delta$ and the ideality factor~$\nid$. The reaction order describes how the recombination rate~$R$ scales with the charge carrier density, $R \propto n^\delta$. For a purely bimolecular process, $\delta = 2$ applies. Conceptually, the ideality factor describes the slope of the exponential dependence of the recombination rate on voltage,
\begin{equation}
R = R_0 \exp\left(\frac{q\voc}{\nid kT}\right),
\label{eq:R_oc}
\end{equation}
where $R_0$ is the recombination rate without photogeneration. Table~\ref{tab:rec} lists typical values of~$\delta$ and~$\nid$ for some relevant recombination processes, namely direct or band-to-band recombination of free carriers, Shockley--Read--Hall recombination via deep traps, and recombination via exponential tails of the density of states.

\begin{table}
\caption{Typical values of the reaction order~$\delta$ and ideality factor~$\nid$ for different recombination mechanisms. Note that the values only apply to the case of balanced electron and hole densities~($n = p$). Also given is the parameter~$\xi$ as it is defined in the text.}
\begin{tabular}{lccc}
\toprule
Recombination mechanism & $\delta$ & $\nid$ & $\xi$\\
 \midrule
Free-carrier & 2 & 1 & 0.5	\\
Shockley--Read--Hall & 1 & 2 & 0 \\
Tail states & $>2$	& 1--2 & 0.5--1	\\
\bottomrule
\end{tabular}
\label{tab:rec}
\end{table}

An alternative representation of the ideality factor is given by 
\begin{equation}
\nid = \frac{q}{kT} \frac{d\voc}{d\ln(\jgen)},
\label{eq:nid}
\end{equation}
where $\jgen = qGL$ is the photogenerated current with $G$ being the spatially averaged generation rate. Equation~(\ref{eq:nid}) is based on the assumption that $R = G$ prevails at $V = \voc$. Given that $G$ is normally proportional to the light intensity, $\nid$ is relatively straightforward to determine from the light-intensity dependence of~$\voc$. This representation of the ideality factor is also called light ideality factor and is more reliable than the determination via dark current--voltage curves, which are influenced by series resistance.\cite{Kirchartz2013}

Compared to the ideality factor, the reaction order is much more difficult to determine. Probably the most common method is to measure pairs of values for the carrier lifetime~$\tau$ and the charge density~$n$ by means of transient photovoltage~(TPV) and charge extraction~(CE) measurements, respectively, and to apply the relation~$R = n/\tau$. However, this method is based on the assumption that during the CE experiment all charge carriers are extracted and that there are no recombination losses. To be independent of this assumption, we used an alternative way to determine~$\delta$, which is based solely on photovoltage measurements and outlined in the following.

Several studies have demonstrated that both the carrier density and carrier lifetime follow an exponential dependence on $\voc$,
\begin{eqnarray}
\label{eq:n}
n = n_0 \exp\left(\frac{q\voc}{\mn kT}\right),\\
\label{eq:tau_n}
\tau = \tau_0 \exp\left(-\frac{q\voc}{\mtau kT}\right),
\end{eqnarray}
where $n_0$ and $\tau_0$ are proportionality constants, while $\mn$ and $\mtau$ determine the slope in a semi-logarithmic representation.\cite{Maurano2011,Foertig2012,Shuttle2008a} The reaction order can be estimated either by plotting $\tau$ against $n$ and fitting the data to $\tau \propto n^{1-\delta}$, or by the slopes of Eqs.~(\ref{eq:n}) and (\ref{eq:tau_n}), 
\begin{equation}
\delta = \frac{\mn}{\mtau} + 1.
\label{eq:delta1}
\end{equation}
We now want to find an alternative for Eq.~(\ref{eq:delta1}) that does not depend on~$\mn$ due to the experimental difficulties mentioned above. We start by dividing Eq.~(\ref{eq:n}) by Eq.~(\ref{eq:tau_n}), which yields the following expression for the recombination rate:
\begin{equation}
R = \frac{n}{\tau}= \frac{n_0}{\tau_0} \exp\left[\frac{q\voc}{(\frac{1}{\mn} + \frac{1}{\mtau}) kT}\right].
\label{eq:R_oc_2}
\end{equation}
Comparison with Eq.~(\ref{eq:R_oc}) yields that $R_0 = n_0/\tau_0$ and gives the following reciprocal expression for the ideality factor:
\begin{equation}
\nid^{-1} = \mn^{-1} + \mtau^{-1}.
\label{eq:ideality_factors}
\end{equation}

Foertig~et~al.\cite{Foertig2012} demonstrated that Eqs.~(\ref{eq:R_oc}), (\ref{eq:nid}) and (\ref{eq:ideality_factors}) are equivalent representations of the ideality factor and that static and transient approaches can provide a consistent picture of the recombination in OPVs. Thus, by plugging Eq.~(\ref{eq:ideality_factors}) into Eq.~(\ref{eq:delta1}), the reaction order~$\delta$ can be determined from any two of the three parameters $\mn$, $\mtau$, and $\nid$. The corresponding relationship that is relevant for this work is
\begin{equation}
\delta = \left(1 - \frac{\nid}{\mtau}\right)^{-1} = (1-\xi)^{-1},
\label{eq:delta2}
\end{equation}
where we have introduced the ratio~$\xi = \nid/\mtau$. Using Eqs.~(\ref{eq:nid}) and (\ref{eq:tau_n}), the parameter~$\xi$ can be written as
\begin{equation}
\xi = \frac{\nid}{\mtau} = \frac{d \voc}{d\ln(\jgen)}\frac{d\ln(\tau)}{d \voc} = \frac{d\ln(\tau)}{d\ln(\jgen)},
\label{eq:xi}
\end{equation}
which implies that the lifetime follows a power law of the form~$\tau \propto G^\xi$. Hence, it is instructive to determine $\xi$ using TPV~measurements, which probe the small-perturbation lifetime~$\taudn$ at various background photogeneration rates~$G$. Using the relationship~$\tau = \delta\taudn$ as demonstrated in Ref.~\citenum{Maurano2011}, we finally arrive at:
\begin{equation}
\taudn \propto (1-\xi)G^\xi.
\label{eq:taudn}
\end{equation}

\begin{figure}[t]
\centering
\includegraphics[width=0.49\textwidth]{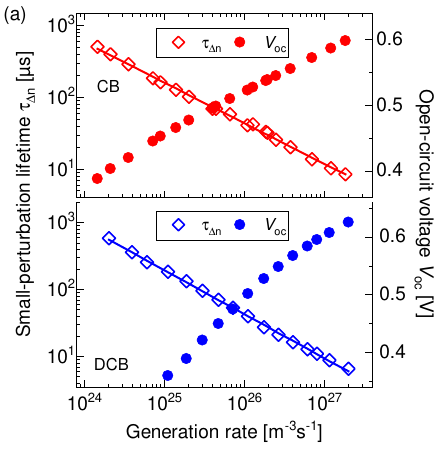}
\hfill
\includegraphics[width=0.49\textwidth]{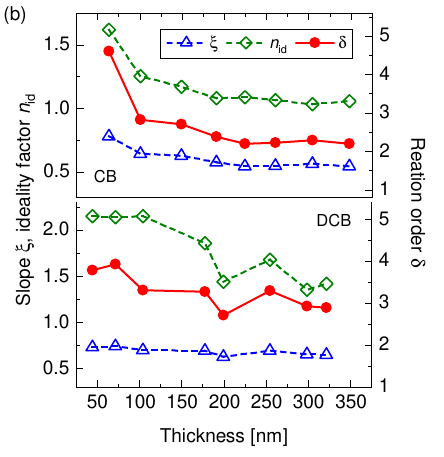}
\caption{(a)~TPV~lifetime~$\taudn$ and open-circuit voltage~$\voc$ as a function of the photogeneration rate for \unit[300]{nm} thick devices. Straight lines are fits to Eq.~{\ref{eq:taudn}}, from which the parameter~$\xi$ was derived. The generation rate was estimated from TMM calculations. (b)~Exponent~$\xi$~(triangles), light ideality factor~$\nid$~(diamonds), and apparent reaction order~$\delta = (1-\xi)^{-1}$~(circles) as a function of the active-layer thickness.}
\label{fig:Rec_1_2}
\end{figure}

Figure~\ref{fig:Rec_1_2}a validates this relationship for both thick CB and DCB devices. Note that at large thickness, the geometrical capacity is low, so that measured lifetimes are supposed to represent the recombination dynamics of photogenerated charges rather than the capacitive discharging of the device.\cite{Kiermasch2018} Hence, fitting the experimental data to Eq.~(\ref{eq:taudn}) and using~$\delta = (1 - \xi)^{-1}$ gives an estimate of the reaction order as long as~$\xi$ is not too close to unity. Figure~\ref{fig:Rec_1_2}a also displays the photogeneration dependence of~$\voc$, determined under the same illumination conditions as the TPV~measurements. While for the CB~device, the whole experimental range is characterized by a single slope, the DCB~device shows a transition towards lower ideality factors at high generations rates~(${\sim}\unit[10^{27}]{m^{-3}s^{-1}}$), roughly corresponding to 1-sun illumination. Hence, the following analysis is limited to medium photogeneration, where the slopes for both kinds of devices are well defined.

Figure~\ref{fig:Rec_1_2}b shows the extracted values of~$\xi$ and~$\delta$ for the complete thickness series, together with the light ideality factor according to Eq.~(\ref{eq:nid}). For the CB~devices, $\delta$ and~$\nid$ exhibit an initial decrease with thickness, but reach fairly constant values of~$\delta = 2.27 \pm 0.08$ and~$\nid = 1.07 \pm 0.02$ between~$L = 150$ and~\unit[350]{nm}. The largely increased reaction order and ideality factor for the 65-nm device may be due to the importance of spatial carrier gradients\cite{Kirchartz2012} or capacitive discharging effects.\cite{Kiermasch2018} In contrast, for the DCB~devices, the trend in thickness is less clear and the data generally show a larger scatter; but it is clearly seen that the reaction order and the ideality factor are generally larger than for the CB devices, ranging from $\delta = 2.7$ to~3.9 and from $\nid = 1.4$ to slightly above~2.

Thus, the DCB~devices clearly show the fingerprints for recombination via exponential tails, while the behavior of the CB~devices is much closer to the expectation for a direct bimolecular recombination mechanism. However, as we discuss in the main text, it is not likely that the DCB~devices have more or deeper tail states. Instead, we assume that direct bimolecular recombination is more reduced than in CB~devices. This assumption is also supported by the fact that the transition of the ideality factor to the bimolecular regime~(see Figure~\ref{fig:Rec_1_2}a) only occurs at much higher generation rates.

\clearpage
\section{Estimation of Recombination Rate Constant}

\begin{figure}[h]
\centering
\includegraphics[width=0.6\linewidth]{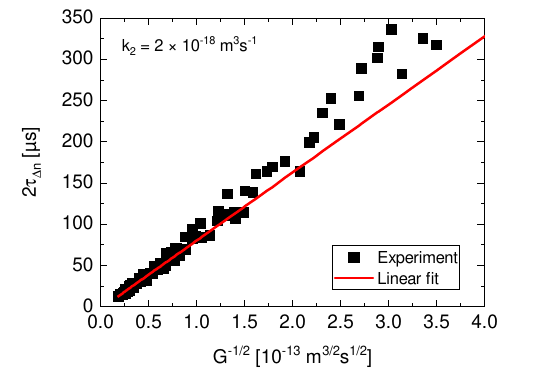}
\caption{Estimation of the recombination rate constant~$k_2$ according to $\tau = (k_2 G)^{-1/2}$~(see main text) for thick CB devices with $L > \unit[150]{nm}$. The total charge carrier lifetime was approximated from the small-perturbation lifetime via $\tau = 2\tau_{\Delta n}$ under the assumption of a strictly bimolecular process. As can be seen, the data points for high generation rates~(i.e., low values of $G^{-1/2}$) collapse into one straight line and a linear fit yields a rate constant of~$k_2 = \unit[2 \times 10^{-18}]{m^3s^{-1}}$.}
\label{fig:tau_vs_genRate}
\end{figure}

\clearpage
\section{Neher Model}

\begin{figure}[h]
\centering
\includegraphics[width=\linewidth]{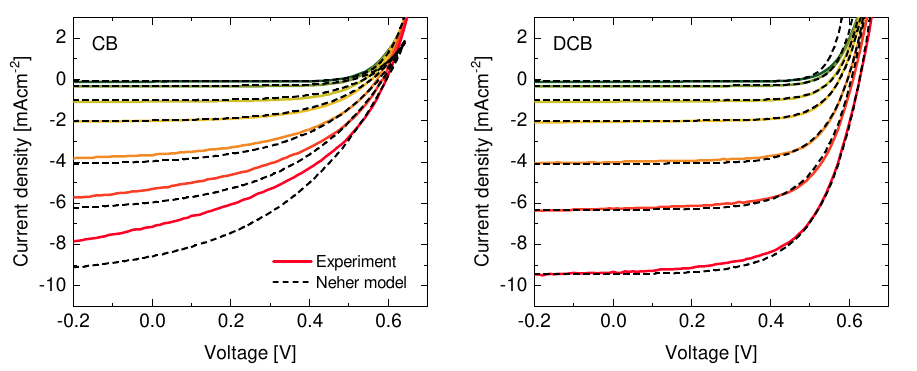}
\caption{Light-intensity dependent $j$--$V$ curves of 300-nm thick devices~(data points) in comparison to the modified Shockley model by Neher~et~al.\cite{Neher2016}~(dashed lines) for recombination rate constants of $k_2 = \unit[2 \times 10^{-12}]{cm^3\,s^{-1}}$~(CB) and $k_2 = \unit[1 \times 10^{-13}]{cm^3\,s^{-1}}$~(DCB). The deviations for the CB~device at high light intensities are due to the build-up of space charge, which is further discussed in the main text.}
\end{figure}

\clearpage
\section{Energy-Level Diagrams}

\begin{figure}
\centering
\includegraphics[width=0.9\textwidth]{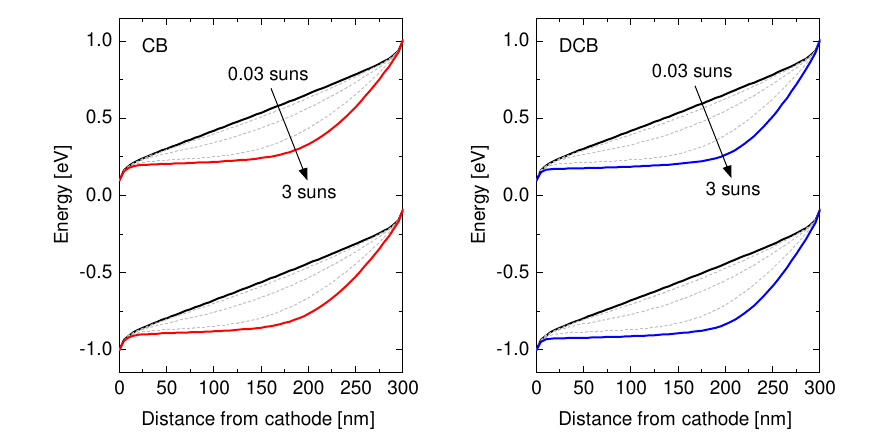}
\caption{Energy-level diagrams under short-circuit conditions for \unit[300]{nm} thick solar cells calculated with a drift--diffusion model.\cite{Burgelman2000,Wilken2020} In both the CB and DCB~device, space charge builds up with increasing light intensity due to the imbalanced charge transport. The width of the space-charge region is about \unit[160]{nm} at 1-sun illumination. Note that the energy levels are independent of the recombination rate constant~$k_2$.}
\label{fig:figureS1}
\end{figure}

\bibliography{supplement}